\documentclass[natbib,smallextended]{svjour3}       
\smartqed  
\usepackage{graphicx}
\usepackage{amsmath, amssymb, txfonts}
\usepackage{mathptmx} 
\usepackage{aps-bibstyle}
\journalname{Space Sci. Rev.}

\usepackage{bm}

\newcommand{\EQ}{\begin{equation}}
\newcommand{\EN}{\end{equation}}
\newcommand{\EQA}{\begin{eqnarray}}
\newcommand{\ENA}{\end{eqnarray}}

\newcommand{\EEq}[1]{Equation~(\ref{#1})}
\newcommand{\Eq}[1]{Eq.~(\ref{#1})}

\newcommand{\Sec}[1]{Sect.~\ref{#1}}

\newcommand{\Fig}[1]{Fig.~\ref{#1}}

\newcommand{\bra}[1]{\langle #1\rangle}

{}
{}
{}

{}
{}
{}
{}
{}
{}
{}
{}
{}
{}
{}
{}
{}
{}
{}
{}
{}
{}

{}
{}
{}

\newcommand{\hatkk}{\hat{\bm{k}}}

{}

{}
{}
{}

\newcommand{\nnn}{\hat{\bm{n}}}

\newcommand{\meanAA}{{\overline{\bm{A}}}}
\newcommand{\meanBB}{{\overline{\bm{B}}}}
\newcommand{\meanJJ}{{\overline{\bm{J}}}}
\newcommand{\meanFF}{{\overline{\bm{F}}}}

\newcommand{\nullvector}{{\bf0}}

\newcommand{\kk}{\bm{k}}

\newcommand{\xx}{\bm{x}}

\newcommand{\aaaa}{\bm{a}}
\newcommand{\bb}{\bm{b}}
\newcommand{\jj}{\bm{j}}
\newcommand{\rr}{\bm{r}}
\newcommand{\grav}{\bm{g}}
\newcommand{\BB}{\bm{B}}
\newcommand{\EE}{\bm{E}}
\newcommand{\JJ}{\bm{J}}
\newcommand{\oo}{\bm{\omega}}
\newcommand{\AAA}{\bm{A}}

\newcommand{\uu}{\bm{u}}

\newcommand{\nab}{{\bm{\nabla}}}

\newcommand{\ii}{{\rm i}}

\newcommand{\curl}{{\rm curl} \, {}}
\newcommand{\dive}{{\rm div}  \, {}}

\newcommand{\dd}{{\rm d} {}}

\def\degr{\hbox{$^\circ$}}

\def\ga{\mathrel{\mathchoice {\vcenter{\offinterlineskip\halign{\hfil
$\displaystyle##$\hfil\cr>\cr\sim\cr}}}
{\vcenter{\offinterlineskip\halign{\hfil$\textstyle##$\hfil\cr>\cr\sim\cr}}}
{\vcenter{\offinterlineskip\halign{\hfil$\scriptstyle##$\hfil\cr>\cr\sim\cr}}}
{\vcenter{\offinterlineskip\halign{\hfil$\scriptscriptstyle##$\hfil\cr>\cr\sim\cr}}}}}
\def\Rm{R_{\rm m}}

\def\alpK{\alpha_{\rm K}}
\def\alpM{\alpha_{\rm M}}

\def\EM{E_{\rm M}}

\def\cs{c_{\rm s}}

\def\kf{k_{\rm f}}

\def\HM{H_{\rm M}}
\def\EM{E_{\rm M}}

\def\urms{u_{\rm rms}}

\def\etat{\eta_{\rm t}}

\def\etaT{\eta_{\rm T}}

\def\Beq{B_{\rm eq}}

\newcommand{\G}{\,{\rm G}}

\newcommand{\kG}{\,{\rm kG}}

\newcommand{\s}{\,{\rm s}}

\newcommand{\km}{\,{\rm km}}

\newcommand{\Mm}{\,{\rm Mm}}
\newcommand{\Mx}{\,{\rm Mx}}

\newcommand{\AU}{\,{\rm AU}}

\begin{document}

\title{The Global Solar Dynamo}


\author{R. H. Cameron    \and
        M. Dikpati    \and
        A. Brandenburg
}


\institute{R. H. Cameron \at
              Max-Planck-Institut f\"ur Sonnensystemforschung,
              Justus-von-Liebig-Weg 3, 37077 G\"ottingen, Germany \\
              \email{cameron@mps.mpg.de}
           \and
           M. Dikpati \at
              High Altitude Observatory, National Center for Atmospheric
              Research 1, 3080 Center Green, Boulder, Colorado 80301
           \and
           A. Brandenburg\at
              Nordita, KTH Royal Institute of Technology and
              Stockholm University, SE-10691 Stockholm, Sweden\\
              Department of Astronomy, Stockholm University,
              SE-10691 Stockholm, Sweden\\
              JILA, Department of Astrophysical and Planetary Sciences, and
              Laboratory for Atmospheric and Space Physics,
              University of Colorado, Boulder, CO 80303, USA
}

\date{Received: date / Accepted: date}

\maketitle

\begin{abstract}
  A brief summary of the various observations and constraints that underlie solar dynamo research
  are presented. The arguments that indicate that the solar dynamo is an alpha-omega dynamo of the
  Babcock-Leighton type are then shortly reviewed. The main open questions that remain are concerned with
  the subsurface dynamics, including why sunspots emerge at preferred latitudes as seen in the familiar
  butterfly wings, why the cycle is about 11 years long, and why the sunspot groups emerge tilted with respect
  to the equator (Joy's law). Next, we turn to magnetic helicity, whose conservation property has been
  identified with the decline of large-scale magnetic fields found in direct numerical simulations at large magnetic
  Reynolds numbers. However, magnetic helicity fluxes through the solar surface can alleviate this problem and connect
  theory with observations, as will be discussed.

\end{abstract}

\section{Introduction}
\label{sec:intro}

The Sun's magnetic field is maintained by its interaction with plasma motions,
i.e. dynamo action. From the 1950s to 1980 the most compelling question of whether dynamo 
action could in fact be responsible for the Sun's magnetic fields 
\citep[a suggestion originally made by][]{Larmor19} was answered. 
That dynamo action is in principle possible for some 
prescribed flows in a body with uniform conductivity was established rigorously in the 1950's 
by \cite{Herzenberg58}. In fact, a number of simple spatially periodic flows were
later shown to act as dynamos \citep{Roberts70}.
From 1955 to about 1980 the next step was taken, where it was shown that
turbulence driven by convection in a rotating system can produce a magnetic field 
with scales comparable to the size of the system \citep{Parker55, Steenbeck66}. The key concept
introduced here is helicity that is the breaking of the symmetry of the convecting 
system by the rotation \cite{Moffatt78,Krause80}. 
This answered the most compelling question of whether dynamo 
action could in fact be responsible for the Sun's magnetic fields,  
as suggested originally made by \cite{Larmor19}. 

The next step was to begin working out which motions are actually producing the 
Sun's magnetic field. Going in this direction substantially beyond what was 
presented by \cite{Parker55}, the work of \cite{Babcock61}, \cite{Leighton64},
and \cite{Leighton69} provide a phenomenological model for the solar dynamo with
close contact to the observations. Their work has been put within the framework
of mean-field models by \cite{Stix74}. Hence the Babcock-Leighton model can
be considered a special example of the mean-field model where the important
dynamo effects are identified with what is seen on the Sun (such as the
``rush to the poles'' and the tilt of bipolar active regions).  
 
The aim of dynamo theory today, in the context of the Sun, is to understand 
how the dynamo actually operates to produce the magnetic fields we see. The
constraints that inform our understanding are thus the various different
observations of solar magnetic fields. We will discuss some of the different
types of observations that are available to us in section 2. We will try to 
present some coverage of what time- and spatial-scales are covered by the observations
but cannot hope to be comprehensive. Some of the observational constraints
will then be presented in their distilled form in section 3 (such as Hale's
law, Joy's law, the Waldmeier effect, some properties of grand minima and maxima).
A few implications for the dynamo that can easily be inferred from the observations 
will be discussed in section 4.

Our coverage of the theoretical and modeling progress that has been made will 
be even more sparse. It is an exciting time when there are a large number
of theoretical models backed by simulations to explain a number of aspects seen on the Sun.
Again our choice of topics and models 
is necessarily biased. In Section 5 we will discuss some prominent results from Flux
Transport Dynamo models, and in Section 6 we will discuss some results from detailed 
calculations of the evolution of the velocity and magnetic fields based on the MHD equations
in a geometry resembling that of the Sun. We comment that because the parameters of the
plasma inside the Sun are extreme, it is beyond the state of the art to hope to realistically
model the solar dynamo at present. These simulations are therefore intended to give
an insight into the physics that could possibly be occurring on the Sun.

\section{What do we know from observations?}

\subsection{What we would like to know?}
Our aim is to understand the solar dynamo. Because understanding requires synthesis,  
even a complete knowledge of the magnetic and velocity fields inside the Sun as a function of 
time would not necessarily be sufficient -- however it would certainly be a 
much better start than what we have at present. What we have is incomplete in many ways.
Firstly we only have reasonably consistent synoptic measurements based on seeing only one half
of the Sun, and only for a few cycles. We know that there are Grand Minima for which we do not have 
this type of data, and in fact the last few cycles were part of a Grand Maximum so we do not have data
representative of even the ``typical'' behavior of the Sun. Finally our only tool for probing the 
convection zone directly is helioseismology which comes with limitations in resolution 
due to finite-wavelength effects and in practice limitations due to the noise associated with granulation. 
Our (partial) blindness to the long-term and interior dynamics is a severe problem for understanding the 
solar dynamo. 

\subsection{What observations do we have?}
Since dynamo action occurs beneath the solar surface, the most directly relevant observations 
for dynamo theory are the time series of surface magnetic and velocity fields. For full-disk almost
continuous space-based observations we have SDO/HMI \citep{Scherrer12} and SOHO/MDI \citep{Scherrer95}
observations. These instruments allow, through inversion of the spectropolarimetric observations, 
the line-of-sight velocity and magnetic field at the solar surface to be inferred (SDO/HMI allows
the full vector magnetic field to be determined). The tangential velocity can be inferred on scales larger
than granules using Local Correlation Tracking techniques \citep[e.g.][]{Roudier98}.  

These observations cover cycle 23 and so far the rising and maximum phase of cycle 24. This allows 
the study of the evolution of the magnetic field on timescales from minutes to a decade or more. Its main 
limitation is that only having two cycles means essentially we only have two data points (on solar-cycle timescales) 
and hence cannot make strong statements about the variation of activity level from cycle to cycle. Also while the resolution 
is adequate for some purposes, it is less than optimal near the poles. 

The limited spatial resolution is partly compensated by high-resolution space missions such as Hinode
\citep{Lites13} that can observe small magnetic flux concentrations near the poles \citep{Tsuneta08}.
The limited number of cycles covered is partly mitigated by observations which decrease in there temporal coverage
and detail as we go further back in time. The most detailed of these are the magnetograms taken in 
synoptic programs which cover most of cycles 21 to 24. These include observations by KPNSO/VTT and SOLIS, 
the Mount Wilson observatory, the Wilcox solar observatory and for the later cycles GONG. These data allow us to 
follow the evolution of magnetic fields on periods of days to years.
 
Still in the era of photographic plates, we have only occasional magnetograms. In their place synoptic programs 
regularly recorded  images of the Sun in white light \citep[for examples see][]{Howard84, Howard90}  
and in some important lines Ca II K \citep[e.g.][]{Bertello10} at Mount Wilson and Kodaikanal.
These types of programs have been undertaken for over 100 years. 

Prior to photography, sunspots were drawn by hand, and we have systematic and continuous records of sunspot
location and areas extending back to 1874 \citep{Balmaceda09}, and less systematically to earlier times
\citep[e.g.][]{Arlt13,Diercke15}. Before this we have sunspot counts going back to before the Maunder Minimum
\citep[see the review by][]{Clette14}.

Only sparse records of direct sunspot observations exist prior to this -- there are for instance occasional 
reports extending back thousands of years. 
Instead we must rely on records of the interaction between the solar magnetic fields with the 
Earth's magnetic field -- these exist in the form of records of auroral \citep[e.g.][]{Kivsky88} and more
systematically in terms of the geomagnetic indices, such as the ``aa'' index. This index can be used to infer the
interplanetary magnetic field near the earth, which in turn can be related to the open flux of the Sun because
the field strength of the radial component of the interplanetary magnetic field largely depends only on the distance
from the Sun \citep{smith95}. The ``aa'' index then gives us a record of global properties of the solar magnetic field
that can be extended back to 1844 \citep{Nevanlinna93}. Nature has also created records that can be used to infer
solar activity, in particular cosmogenic nucleotides stored in ice cores, which extends our knowledge back to over
9600 years \citep{Steinhilber12}.

\subsection{Constraints}

The dynamo problem is essentially concerned with plasma motions generating and sustaining magnetic fields.
We therefore begin by {\em{extremely briefly}} outlining what is known about the motions themselves.

\subsubsection{The flows}

\paragraph{Granulation and supergranulation.}

Heat is transported by convective motions in the outer 30\% of the Sun. The dominant scale of the
convection near the surface (granulation) is well understood and depends mainly on the Sun's mass
and luminosity \citep{Stein89}. The properties of flows at larger scales
(both supergranulation and the lack of giant cells) are much more poorly understood theoretically 
\citep{Lord14} and observationally (compare \citeauthor{Hanasoge12}, \citeyear{Hanasoge12}
and \citeauthor{Greer15}, \citeyear{Greer15}).

The interaction between the convection and rotation, especially at larger scales, drives global-scale
flows such as differential rotation and meridional circulation. Improving our understanding of the large-scale
convection is therefore a priority.

\paragraph{Rotation.}

The total angular momentum of the Sun is a result of the angular momentum it had when it formed,
and its evolution since then (e.g. due to magnetic breaking). The total angular momentum of the
Sun, and in particular of the convection zone, is thus a basic parameter from the point of
view of dynamo theory.

\paragraph{Differential rotation -- Latitudinal and radial.}

The Sun's differential rotation is well known as a function of both latitude and radius \citep{Schou98}.
The main properties are that latitudinal shear is much greater that the radial shear, the latter of
which is localized at the tachocline at the base of the convection zone and in a near-surface shear layer.
\citep[For a detailed review see][]{Howe09}.

\paragraph{Torsional oscillations.}

The time dependence of the differential rotation is called ``torsional oscillations'' \citep{Howard80, Schou98}.
These are clearly associated with magnetic activity, but are probably too weak to significantly influence the
evolution of the Sun's magnetic field (they are however likely to be an important diagnostic).

\paragraph{Meridional flow.}

There is also a large-scale meridional flow, with a well-observed poleward component at the solar surface
\citep{Duvall79, Ulrich10}. 
The subsurface structure of the flow is more controversial \citep[compare][]{Zhao13, Schad13, Jackiewicz15}.
Given the important role of the subsurface meridional flow in transporting the field in the
Flux Transport Dynamo model (discussed in Section~4.1) resolving this controversy should be
seen as a priority.

\subsubsection{Magnetic field evolution}

\begin{figure}[h!]\begin{center}
\includegraphics[height=.33\columnwidth]{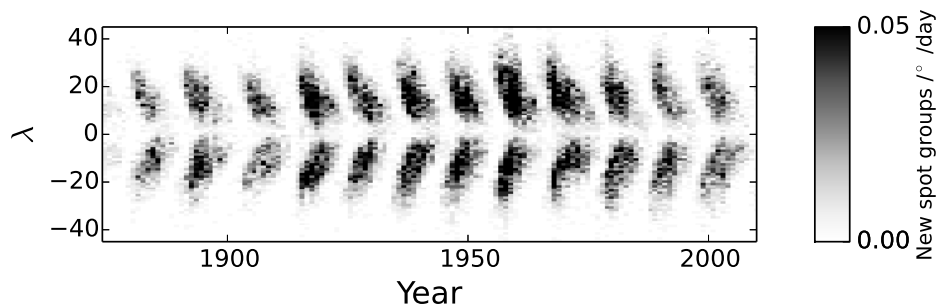}
  \end{center}
  \caption{A butterfly diagram: the number of sunspots appearing as a function of latitude $\lambda$ and time
          (based on Royal Greenwich Observatory and  USAF/NOAA SOON data).}
\label{fig:bf}
\end{figure}

\begin{figure}[h!]\begin{center}
\includegraphics[height=.5\columnwidth]{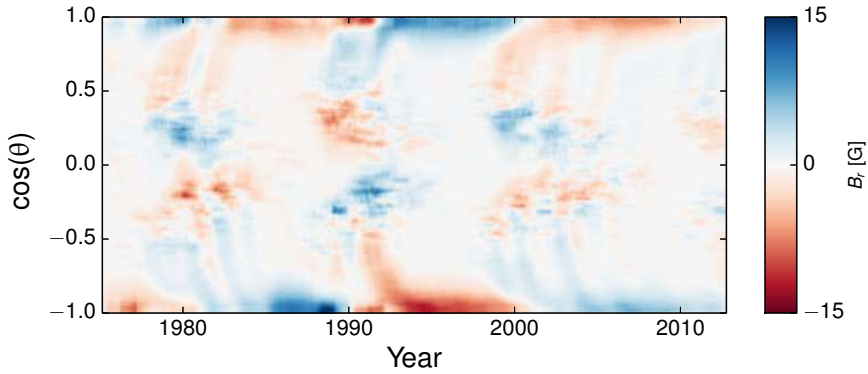}
  \end{center}
  \caption{A magnetic butterfly diagram: the longitudinally averaged radial component of the solar magnetic field
           as a function of latitude and time (based on Kitt Peak National Observatory Synoptic magnetograms).}
\label{fig:mbf}
\end{figure}

Observations of the type described in Section~2.2 are the constraints we have for the solar dynamo.
They contain a lot of information, some of which can be summarized in simple figures, and some
of which can be distilled into ``laws''. Two figures which contain a lot of information are
the butterfly diagram \citep[][see also Figure~\ref{fig:bf}]{Maunder1904} and the magnetic butterfly map
(Figure~\ref{fig:mbf}).  

From the observations a number of properties (some of them ``laws'') have been described in the literature.
A comprehensive solar dynamo model should be able consistent with these observational constraints. 
We begin with a number of constraints that are directly related to the magnetic activity.

\paragraph{11 year activity cycles, 22 year magnetic cycles.}

A good proxy of magnetic activity is the number of sunspots, which varies in time
with minima every 10 to 12 years. This was first noted by \cite{Schwabe1849}. Different 
cycles have different amplitudes \cite[and modulation on longer periods might be present, 
e.g.][]{Gleissberg39}.
The 11 year solar activity cycle corresponds to half of the 22 year magnetic cycle
\citep{Hale19}, with the dominant polarity of the leading sunspots in each hemisphere 
changing between each activity cycle.

\paragraph{Sp\"orer's Law.}

The emergence location of sunspots is observed to migrate equatorward during the cycle
\citep{Carrington58, Spoerer79}, beginning at about 35$^{\circ}$ and propagating to about
8$^{\circ}$ at the end of the cycle. The rate of propagation is similar for all cycles 
\citep{Waldmeier55, Hathaway11b}.

\paragraph{Hales law.}

The magnetic nature of sunspots was discovered by \cite{Hale08}. Sunspots typically
appear in groups, with the leading and trailing spots (with respect to the solar rotation)
having different polarities. The leading spots in each hemisphere mostly have the same polarity,
and the polarity is opposite in the other hemisphere. The polarities of the leading and following spots
switch between cycles \citep{Hale19}. 

\paragraph{Joy's law.}

As implied by Hales Law, sunspots often appear as bipolar pairs with the leading spots during one cycle
and in one hemisphere having the same polarity. This is a statement about the  east-west orientation of the
sunspots. There is also a tendency for the leading spots to be slightly closer to the equator than the
following spots. This tendency is much weaker than that of Hale's law, with the angle implied by the
North-South separation compared to the East-West separation being about 7 degrees.  This effect is known as Joy's
law and was reported by \cite{Hale19}. There is some evidence that the strength of the effect depends on
the strength of the cycle \citep{Dasi10}.

The effect is also much
weaker in the sense that there is a lot of scatter in the North-South separation so that the effect
is only robust when a large sample of sunspots is considered.

\paragraph{Waldmeier effect.}

The Waldmeier effect states that strong cycles peak earlier than weak cycles \citep{Waldmeier41}
\citep[although the Waldmeier effect does not appear in all measures of the solar activity, cf][]{Dikpati08, Cameron08}.
There is a closely related fact that strong cycles rise quickly, which \cite{Karak11} call WE2, for the
Waldmeier effect 2.

\paragraph{North-South asymmetry.}

Cycles are not symmetric \citep{Spoerer89a}, and interestingly the asymmetric behavior can be coherent over
many cycles \citep{Carbonell93}.

\paragraph{Extended cycle.}
While the sunspot number has a period of around 11 years, the butterfly diagram indicates that the wings overlap
so that sunspots corresponding to each cycle are present for about 13 years. Smaller than sunspots, ephemeral regions
associated with a cycle have been shown to emerge about 5 years earlier, so the activity related to one cycle
extends to about 18 years \citep{Wilson88}.

\paragraph{Correlation between polar fields, open flux and strength of next cycle.}

There is a strong correlation between the polar field at minimum (determined using polar faculae as a proxy)
and the strength of the next cycle \citep{Munoz-Jaramillo13}; a stronger correlation exists between the Sun's open
flux, determined using the minima of the $aa$ index as a proxy, and the strength of the next cycle \citep{Wang09}.
\cite{Cameron07} have suggested that this might be accounted for by the overlapping of cycles combined with the
Waldmeier effect. A commonly claimed effect that the length of a minimum correlates positively with the weakness of
the next cycle peak has been shown from sunspot data by \citep{Dikpati10a} to be false for the most recent 12 cycles.'

\paragraph{Magnetic fields at the surface are advected by surface flows as if they were corks.}

Outside of active regions, the radial component of the magnetic field is advected by the horizontal
component of the velocity as if it were a passive tracer \citep{DeVore84}. The details of the modeling and
observations that support this were reviewed recently in \cite{Jiang14}.

\paragraph{Coronal Mass Ejections and  magnetic helicity fluxes.}

The structure of the magnetic field in the solar atmosphere, with filaments and sigmoid shaped active regions
as well as the various types of activity in the solar atmosphere, such as flares and coronal mass ejections
also contain information related to the solar cycle. The interpretation of these structures in terms of
the helicity generated by the dynamo is however not straightforward \citep{Zirker97}. We return to the helicity
in some detail in Section 5.

\paragraph{Grand minima and maxima.}

The above ``laws'' and properties of the magnetic activity are mainly based on the last few hundred years of data.
We however know that the solar dynamo does not always behave like this -- there are extended periods of
low activity including the Maunder minimum \citep{Spoerer89}. They occur on average every 300 years
\citep{Usoskin07} and presumably represent a different state of the dynamo. We have very few observational
constraints for this state of the dynamo and therefore do not discuss it further in this paper.

\section{Synthesising the observations and theory}
We now turn to the open problem of synthesizing the observations and the ``laws'' they embody with
well known basic physics in order to gain an understanding of the solar dynamo. 

\subsection{The omega and alpha effects}

\cite{Cameron15} used the simplicity of the toroidal field implied by Hale's law to show that
the surface magnetic field plays a key role in the solar dynamo. The argument begins by
noting that Hale's law tells us that in each hemisphere, and during one cycle, the leading spots mainly
\citep[about 96\% for large active regions,][]{Wang89} have the same magnetic polarity.
This strong preference for the leading spots to have the same polarity indicates that the
spots are coming from the emergence of toroidal flux that is all of the same polarity.
The authors then considered the induction equation 
\EQ
   {\partial\BB\over\partial t}=\nab\times(\uu\times\BB-\eta\mu_0\JJ),
   \label{eqn:induction}
\EN
where $\uu$ and $\BB$ are velocity and magnetic fields, $t$ is time,
$\JJ=\nab\times\BB/\mu_0$ is the current density,
$\mu_0$ is the magnetic permeability, and $\eta=1/\sigma\mu_0$ is the
magnetic diffusivity with $\sigma$ being the conductivity.
They then applied Stokes theorem to with a contour in a meridional plane and encompassing the
convection zone in the northern hemisphere. This allowed them to demonstrate that the
generation of net toroidal flux in each hemisphere is dominated by the winding up of the
poloidal flux threading the poles of the photosphere at the poles by the latitudinal differential
rotation.

The polar fields themselves are the remnants of flux that has crossed the equator \citep{Durrant04},
which is dominated by either the emergence of tilted active regions across the equator \citep{Cameron13} or
the advection of active region flux across the equator due to the random shuffling of the field lines
due to the supergranular flows.

The simplicity of the toroidal magnetic field at solar maxima, and the poloidal magnetic field at solar minima,
therefore indicates that the solar dynamo is of the Babcock-Leighton type. Explicitly it is an
alpha-omega dynamo where the relevant poloidal field threads through the photosphere. The omega effect is
simply the winding up of this poloidal field by the latitudinal differential rotation. The alpha effect is
what produces the tilt of the active regions with respect to the equator (Joy's law).

We comment that the question of why we have a butterfly diagram, why sunspots obey Joy's law, why the cycle
length is around 11 years, and why sunspots only emerge below about 40 degrees all remain open questions.
They are difficult to answer because they are intimately related to the subsurface dynamics, which are
mostly poorly understood theoretically and observationally. In the next two subsections we discuss some of the ideas
which are in the literature.

\subsection{Equatorial migration of the butterfly wings.}
\begin{figure}[h!]\begin{center}
\includegraphics[width=1.0\textwidth]{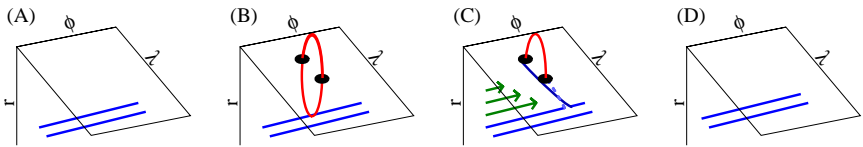}
  \end{center}\caption[]{Illustration of the cause of the equatorial propagation in the dynamo
    wave model. The model begins with toroidal field (shown in blue) situated away from the equator
    in Panel A -- the equator is towards the top of the panel. In Panel B a  sunspot group emerges
    from the toroidal flux. The rising tube is acted on by the Coriolis force, and by the convection,
    and poloidal field is produced (this is the alpha effect). In Panel C the poloidal field is sheared
    by radial differential rotation, producing positive and negative toroidal field. The dashed blue
    line represents toroidal field of the opposite sign to the original (solid) toroidal field.
    In Panel D the dashed poloidal field has canceled the existing toroidal field at high latitudes. The
    low lying newly created toroidal field leads to a new band of toroidal field with the same orientation as
    the existing field but at lower latitudes. The net action has thus been to move the toroidal field closer
    to the equator.}
  \label{fig:RHC_DW}
\end{figure}

Sp\"orer's law (discussed in Section~2.3.2) states that the latitude of emergence of sunspots propagates towards
the equator as the cycle proceeds. The most straightforward (and probably correct) interpretation of this is that
the underlying toroidal field is propagating towards the equator. There have been two main suggestions to explain
the propagation of the toroidal field. The first is that of \cite{Parker55}, and explains the equatorward propagation in
terms of a dynamo wave. The cause of the equatorial propagation in this model is explained in Figure~\ref{fig:RHC_DW}
in the case of a Babcock-Leighton dynamo.
The essential idea is that radial differential shear causes toroidal flux to propagate latitudinally. The direction
of propagation (equatorwards or polewards) depends on the sign of the alpha effect (which generates poloidal flux from
toroidal flux)  and the on whether the differential rotation rate increases or decreases with depth \citep{Yos75}.

In terms of the Sun, the sign of the alpha effect is generally believed to be such that differential rotation has to
increase with depth to obtain equatorward propagation. This is an issue on the Sun where the differential rotation
decreases inwards near the base of the convection zone in the range of latitudes where sunspots form \citep{BM87}.
Thus if the solar dynamo is substantially located in the tachocline then this mechanism is excluded, the mechanism is however
viable if the winding up of the toroidal field occurs in the near-surface shear layer.

\begin{figure}[h!]\begin{center}
\includegraphics[width=1.0\textwidth]{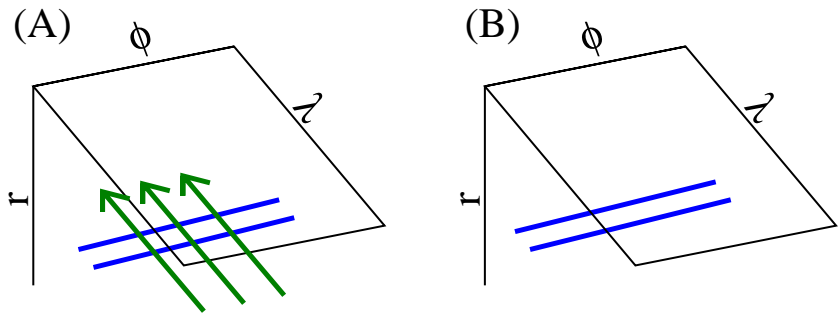}
  \end{center}\caption[]{Illustration of the cause of the equatorial propagation in the flux transport dynamo model --
    the toroidal flux (blue lines) is simply advected towards the equator by the meridional circulation (green arrows).
    The question are whether such a flow exists, where the toroidal field
    is stored and whether diffusive transport plays a role.}
  \label{fig:RHC_FTD}
\end{figure}

The second mechanism for explaining the equatorward propagation is shown in Figure~\ref{fig:RHC_FTD} and is supposed to
work at the base of the convection zone. The essential idea is that, if the meridional flow near the base of the tachocline
is sufficiently strong, and if the latitudinal transport due to diffusion is sufficiently weak, then the simple advection
of the toroidal flux by the meridional flow can produce the observed equatorial migration
\citep{Wang91, Choudhuri95, Durney95, Dikpati99, Kueker01}.

Which, if either, of these two explanations accounts for the transport on the Sun remains an open question.

\subsection{Emergence latitudes}
A second important question, which is probably closely related, is why sunspots overwhelmingly appear at latitudes
below 40 degrees in latitude (see Figure~\ref{fig:bf}). An obvious explanation for this would be that the toroidal
flux is concentrated at low latitudes, however this explanation is problematic for Babcock-Leighton type dynamos where
the latitudinal differential shear occurs over essentially all latitudes. A second possible mechanism considers the
instability which leads to flux emergence.  This explanation assumes that the toroidal flux is stored in or near
the tachocline. In this case flux tubes must have a field strength of about $10^5$~G in order to have a substantial
number of emergences at low latitude \citep{Choudhuri87}. The argument is that if the tubes are weaker then their
rise is more affected by the Coriolis force which preferentially causes them to rise with at a constant distance
from the rotation axis (i.e. along cylinders). The value of $10^5$~G is nicely consistent with the field strength
required to produce Joy's law \citep{DSilva93}. Such flux tubes turn out to be much more unstable \citep{Parker55b}
at low latitudes than at high latitudes  \citep{Caligari95}, hence if the toroidal flux is stored in the tachocline
then even if the $10^5$~G loops form at high latitudes the loops are unable to escape to the surface.

This is an attractive possibility because the value of $10^5$~G links several observational results.
However we caution that this effort was focussed on the case where the storage is in the tachocline or near the
base of the convection zone because this was the dominant paradigm at the time the work was carried out. Much
less work has been put into considering why sunspots don't emerge at high latitudes in the case where the toroidal
field is stored in the bulk of the convection zone. For this reason we regard the issue of the latitudinal range
of the butterfly diagram as an open question.

\subsection{Length of the solar cycle}
It is somewhat sobering that after more than 160 years after the discovery by \cite{Schwabe1849} that the
level of solar activity varies with a period of about 11 years that we still don't have a good idea of why it is
11 years.

In principle the length of the solar cycle should reflect the latitude range over which activity appears (say from
a latitude of 30 degrees at the start of a cycle to 8 degrees at the end of a cycle) and the rate at which it
propagates towards the equator. Within the framework of flux transport dynamos, the rate at which activity
propagates towards the equator and the period of the dynamo, is largely determined by the meridional flow circulation
rate \citep{Dikpati99}, however we have no clear basis in either observations or theory for understanding why the
strength of the meridional flow at depth should be such that the dynamo period is 11 years. Within the alternative frame of
a dynamo wave framework the period is set mainly by the magnitude of the alpha and omega
effects, and the alpha effect in particular is not well understood.

To end this section on a positive light,  it is clear that improving our understanding both what causes the equatorward propagation
and what sets the latitudinal extent of the butterfly diagram, will give us a clearer handle on why the period of the solar cycle
is 11 years.

\subsection{The alpha effect}

\label{SurfaceHelicities}
The above discussion indicates that the net toroidal field in each hemisphere is 
produced by the winding up of poloidal field by latitudinal differential rotation.
In this regard it is only the field that threads through the surface which has an
effect on the net flux. This winding up of the field is the omega effect of an 
alpha-omega dynamo. The details of how the alpha effect actually works in the Sun is 
less well understood -- we observe the emergence of magnetic bipolar regions which are 
systematically tilted with respect to the equator, we observe their subsequent evolution,
and we can infer (as in the previous section) that this is the field which is the poloidal
field which gets wound up to produce the net toroidal flux in each hemisphere. What we do not 
observe are the processes which cause the field to emerge with a tilt.

It is relatively clear that the Coriolis force is implicated in the tilt; the uncertainty is whether
the Coriolis for acts directly on the flows associated with the rise of the flux tube \citep[e.g.][]{DSilva93},
or whether it acts on the convective flows which then interacts with the flux tube \citep{Parker55}. This question
is currently completely open.

Even without a proper understanding of the subsurface processes we can include flux emergence and
Joy's law into idealized mean-field simulations. Most of the recent work along these lines has
been in the context of the flux transport dynamo, and the next section will outline some results from
these efforts.

\section{Modeling dynamo action}

\subsection{Flux-transport dynamos}

Two-dimensional ``flux-transport'' dynamos incorporate in some form all of the
processes discussed above, including the idea of flux emergence from toroidal
flux at the base of the convection zone producing tilted active regions at the
surface. These models have been successful in simulating the most important features of the solar 
cycle, including the butterfly diagram, polar field reversals near solar cycle 
maximum, certain global coronal features, and certain asymmetries between North 
and South hemispheres.

\begin{figure}
\includegraphics[width=0.35\textwidth]{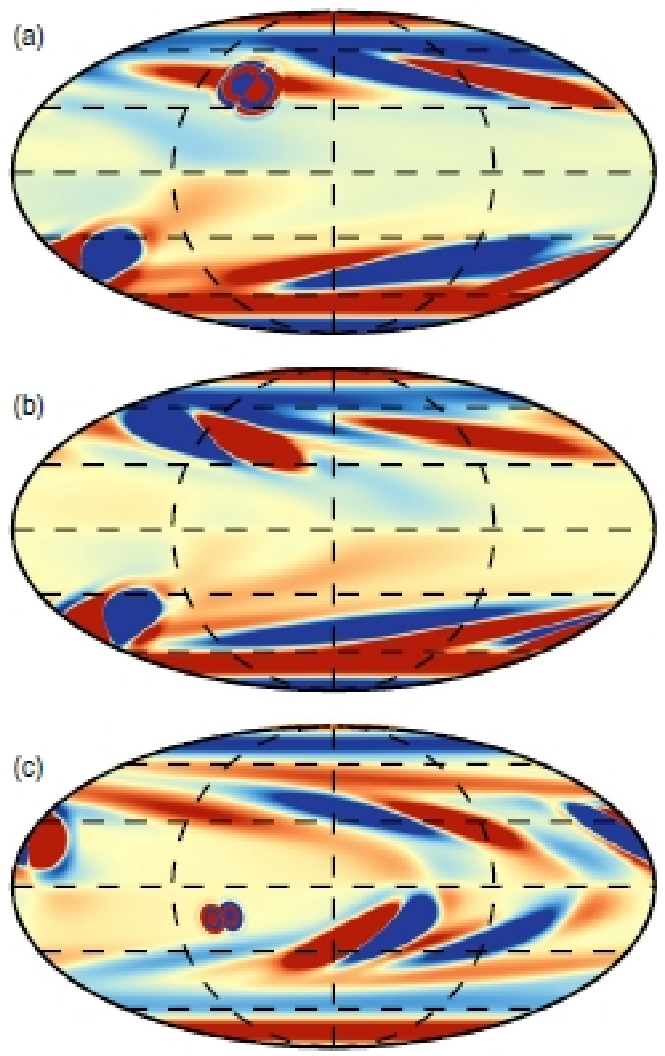}
\includegraphics[width=0.55\textwidth]{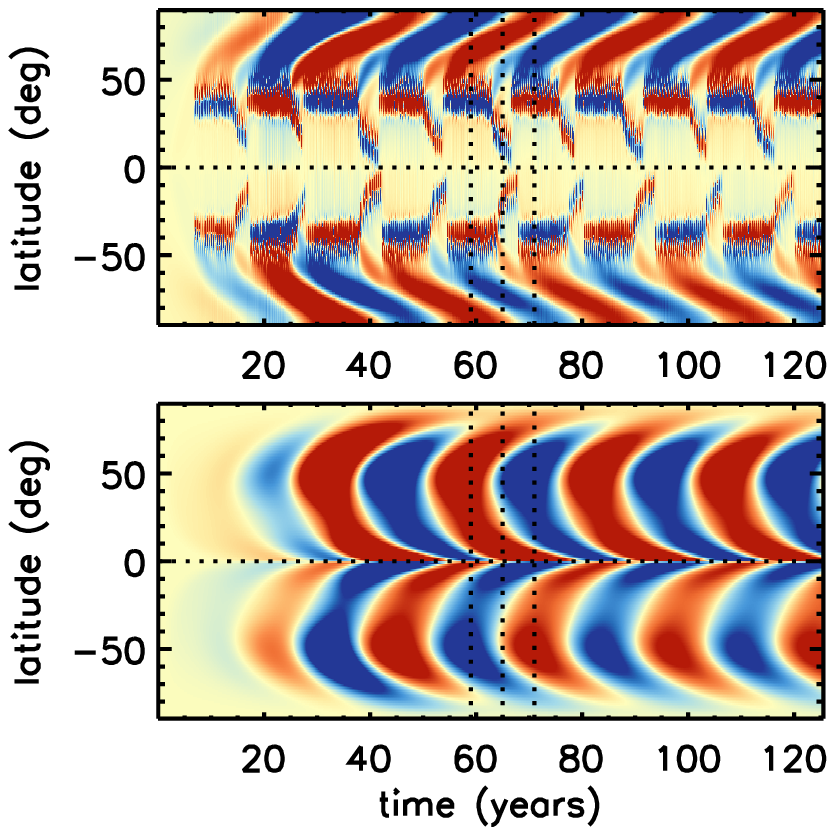}
\caption{Left frame: Eruption of tilted, bipolar spots and their
dispersal by diffusion, meridional circulation and differential rotation;
right frame: drift of trailing flux towards the poles appears in a
series of streams, and cause polar reversal (top right) and toroidal field
butterfly diagram (bottom right), showing equatorward migration and
dynamo cycle-period governed by meridional circulation speed (frames adopted
from Miesch \& Dikpati (2014)).}
\label{3DBL}
\end{figure}

These models were used to simulate and predict the timing and
amplitude of solar cycle 24 \citep{Dikpati06, Choudhuri07, Dikpati10, Nandy11},
with limited success.
There are several possibilities for this. Some of the suggestions of what was not included (but needs to be) are
changes in the global meridional circulation profile and speed \citep{Belucz13},
or localized inflow cells associated with active regions \citep{Cameron12, Shetye15}
or the scatter in tilt angles \citep{Jiang15}.
Unfortunately the results from helioseismology are too divergent to provide guidance on changes in the meridional flow
\citep[for example, compare the results in recent publications][]{Ulrich10, Zhao13, Schad13, Jackiewicz15}.

One possibility, supported by the observations, is that the weak cycle 24 is the result of the actual
values of the tilt angles in cycle 23 \citep{Jiang15}. The essential point here is that poloidal source
term in the Babcock-Leighton model is based on Joy's law, which is extremely noisy. The noise in Joy's law
translates to noise in the alpha effect, and thus to the strength of the different cycles. For cycle 23
\cite{Jiang15} used the observed tilt angles \citep{Li12} and showed that alpha effect was indeed weak for cycle 23
(thus accounting for the weak cycle 24).

Current work includes extending the simulations to 3D, see, e.g., \cite{Miesch14}. An 
example of such a simulation is shown in Figure~\ref{3DBL}, which depicts the 
longitude-latitude pattern of emerged flux, and the patterns of surface mean 
poloidal fields and deep-seated toroidal fields that are created in a sequence 
of solar cycles.

However, perhaps the biggest challenge to the flux-transport dynamos model is whether 
the meridional circulation is single-celled or multiple-celled or is indeed
not steady at all.
Many solar cycle features can be reproduced well by a flux-transport dynamo
model if the meridional circulation contains a single cell in each hemisphere,
but models with two cells in depth do not \citep{Belucz15}.
There is currently no consensus from observational evidence and 
meridional circulation models on the dominant profile of meridional flow 
in depth and latitude within the convection zone and tachocline. At the 
photosphere there is clearly one primary, poleward cell or often along with
that a weak reverse cell appears at high latitudes. If it is established 
that there are two cells in depth, then another paradigm shift in solar 
dynamo theory will be needed; however it seems to be too early to decide on
that issue now.

\subsection{Global convective dynamo simulations}
\label{Global}

Attempts have also been made to simulate the convective motions and dynamo
action in the Sun, with as few assumptions as possible -- essentially from first principles.
The first semi-successful attempts go back to the early 1980s when
two different groups used the Cray~1 computer at the National Center
for Atmospheric Research in Boulder (Colorado):
Peter Gilman using a model of rotating convection \citep{GM81}
and \cite{MFP81} using forced turbulence in a Cartesian domain.
Both studies provided remarkable first steps into numerical studies
of large-scale and small-scale dynamos, but they also demonstrated that
simulating the Sun will be difficult.

Subsequently, \cite{Gil83} obtained cyclic solutions that were however quite
different from the Sun: instead of equatorward migration of magnetic activity,
he obtained poleward migration.
Furthermore, small-scale dynamo action was not well understood at the
time and the original paper by \cite{MFP81} did not even quote the
now important reference to the paper by \cite{Kaz68}.
The role of helicity was also not clear in some of those first studies,
see also \cite{KYM91}.
This was partly because those early simulations did not have sufficient
scale separation, i.e., the scale separation ratio $\kf/k_1$, where
$k_1$ is the smallest wavenumber in the domain, was too small, 
as was noted subsequently \citep{HBD04}.
In agreement with earlier work on rapidly rotating convection \citep{Gil77},
the contours of angular velocity tend to lie on cylinders.
This implies that the radial gradient of the local angular velocity,
$\partial\Omega/\partial r$, is positive.
Therefore, as expected predictions from mean-field dynamo theory, the
dynamo simulations of \cite{GM81} and \cite{Gla85} produced poleward
migration, and were thus unable to reproduce the solar butterfly diagram.
\citep[These simulations were subsequently applied to the geodynamo problem by
][where success was much clearer.]{GR95}

From these simulations it is clear that reproducing the basic properties of the
convection and the large-scale flows it drives are essential for a complete understanding
of why a rotating middle-aged star such as the Sun should have an 11 year activity cycle
and a butterfly diagram with equatorward propagation.

Computing power has increased dramatically since 1981, and
simulations in more turbulent regimes became possible. This leads to
flow patterns departing from otherwise nearly perfectly cylindrical contours
\citep{METCGG00}.
Global simulations are now being conducted by many groups.
Simulations with the anelastic spherical harmonic (ASH) code \citep{BMT04}
work with a mean stratification close to that of mixing length theory,
but at the solar rotation rate the resulting dynamo is statistically steady.
Only at higher rotation rate, the solutions become time-dependent and cyclic
\citep{BBBMT10}.
Simulations with the EULAG code \citep{GCS10,RCGBS11} also produce cyclic
solutions, although the latitudinal migration of the mean magnetic field
is weak.
The pattern of meridional circulation in the simulations is, to date,
mostly multicellular. This is in stark contrast to mean-field (flow) models,
in which differential rotation is produced by the $\Lambda$ effect and meridional
circulation is dominated by one large cell \citep{BMT92}. 
As mentioned above, helioseismology does not yet provide a consistent answer to
guide the theory or simulations.

A (surprising) key issue which has emerged is the convective power spectrum.
Observations through correlation tracking \citep{RMRRBP08,Hat12}
and helioseismology \citep{Hanasoge12} suggest that the power in
large-scale convective flows (giant cells) is very small.
As alluded to in the introduction, the helioseismic evidence is contested
\citep{Greer15}, so the question remains open on the observational side.
As already suggested by \cite{Spr97}, the structure of the large-scale convection is
also unclear  on the theoretical side -- global simulations miss the
physics of the radiating surface, and this may turn out to be crucial
for producing simulations with realistic flow structures, which may be
dominated by what is known as `entropy rain'.
This would be a fast small-scale downflow originating from the surface
in such a way that the bulk of the convection zone is nearly isentropic,
but with a slightly stable stratification so as not to produce giant cell
convection in the deeper parts \citep{Bra15}.
Clearly, simulations must eventually be able to reproduce the Sun,
and reproducing the convective power spectrum and the large-scale flows is
probably a precondition to accurately reproducing the solar dynamo. Knowing what those
flows are is essential.

Currently the convective simulations are most useful in providing guidance on
what is possible, and elucidating mechanisms. For example in simulations by \cite{KMB12} using the
{\sc Pencil Code}\footnote{https://github.com/pencil-code. }, there was
pronounced equatorward migration, which was later identified as being due
to a local negative radial $\Omega$ gradient \citep{WKKB14}.
While this is a feature not expected to be present in the Sun, it does
demonstrate that the dynamo wave in global simulations follows closely
that expected from mean-field simulations \citep{Parker55,Yos75}.
Global simulations have also produced evidence for strong ($\approx40\kG$)
flux tubes in the bulk of the convection zone \citep{NBBMT14,FF14}.
This is interesting in view of understanding the overall magnetic flux
concentration required to form active regions, although further
amplification is needed; see \cite{SN12} and \cite{MBKR14} for
possible mechanisms.

Both flux transport and full 3D dynamo models necessarily contain parametrizations of processes acting
on scales smaller than spatially resolved. These parametrizations are all rooted in formulations of MHD
turbulence, in which helicity of velocities, magnetic fields and electric currents play central roles.
Therefore we consider helicity effects in detail in section 5

\section{The roles of magnetic helicity}
As revealed above, we are still struggling to understand the true nature of the solar dynamo
and the relation between the large-scale dynamo in the Sun and in global simulations.
In this regard we need to be sure that what we see in the simulations survives in the limit of
large magnetic Reynolds numbers such as are found in the Sun. This has not been yet been fully
confirmed. One example of this is that in the simulations the large-scale field magnetic field
(as opposed to the small-field magnetic field) is often found to decrease with increasing
values of $\Rm$. In this section we show that this can be understood quantitatively in terms of
magnetic helicity conservation, as will be discussed below.

Magnetic helicity is a conserved quantity in ideal MHD, and as such is conserved in the
absence of microphysical diffusivity (i.e. it is not changed by the turbulent flow, or the
associated turbulent magnetic diffusivity). In a closed volume, at low microphysical diffusivity
of the sun, the amount of helicity is not expected to change on timescales of the solar cycle.
This is remarkable and unprecedented in hydrodynamic turbulence.
The only way that magnetic helicity can change and evolve (e.g., over
the course of the 11 year cycle) is through magnetic helicity fluxes.
They can be determined at the surface and this allows contact to be made
between theory and observations. In this section we will therefore outline some of the
different attempts to determine the helicity flux through the solar surface and what this tells
us about the solar dynamo.

There are other types of helicity, including kinetic helicity, which also play important roles in dynamo theory.
Most importantly kinetic helicity makes it relatively simple to generate a large-scale magnetic field \citep{Mof69}.
We therefore begin by discussing the different types of helicity and their definitions. 

\subsection{Definitions.}

Mathematically, helicities are so-called pseudo-scalars, i.e., they are
the dot product of a proper vector and an axial vector.
(The latter changes its orientation when viewed in a mirror.)

\begin{figure}[h!]\begin{center}
\includegraphics[height=.25\columnwidth]{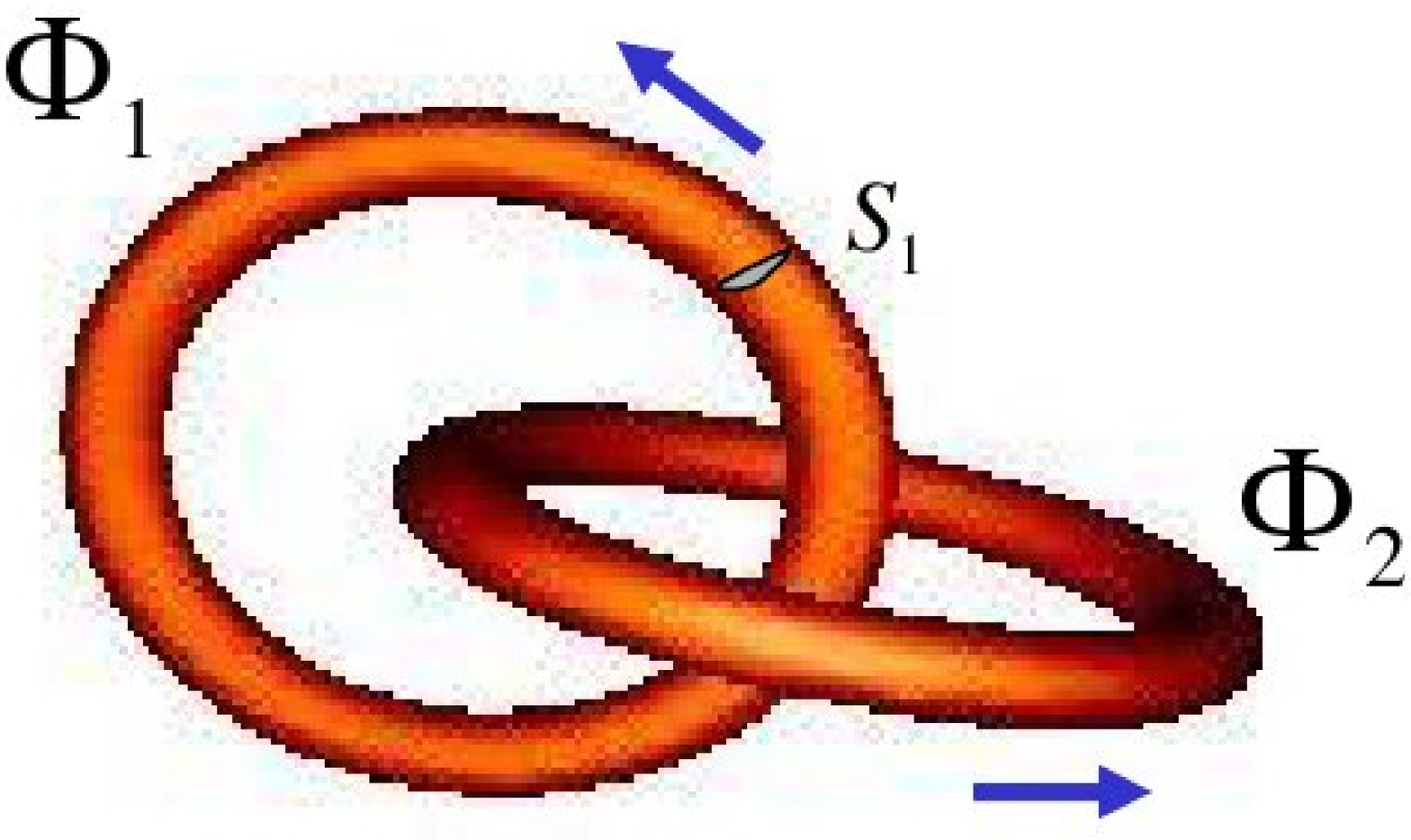}
\includegraphics[height=.25\columnwidth]{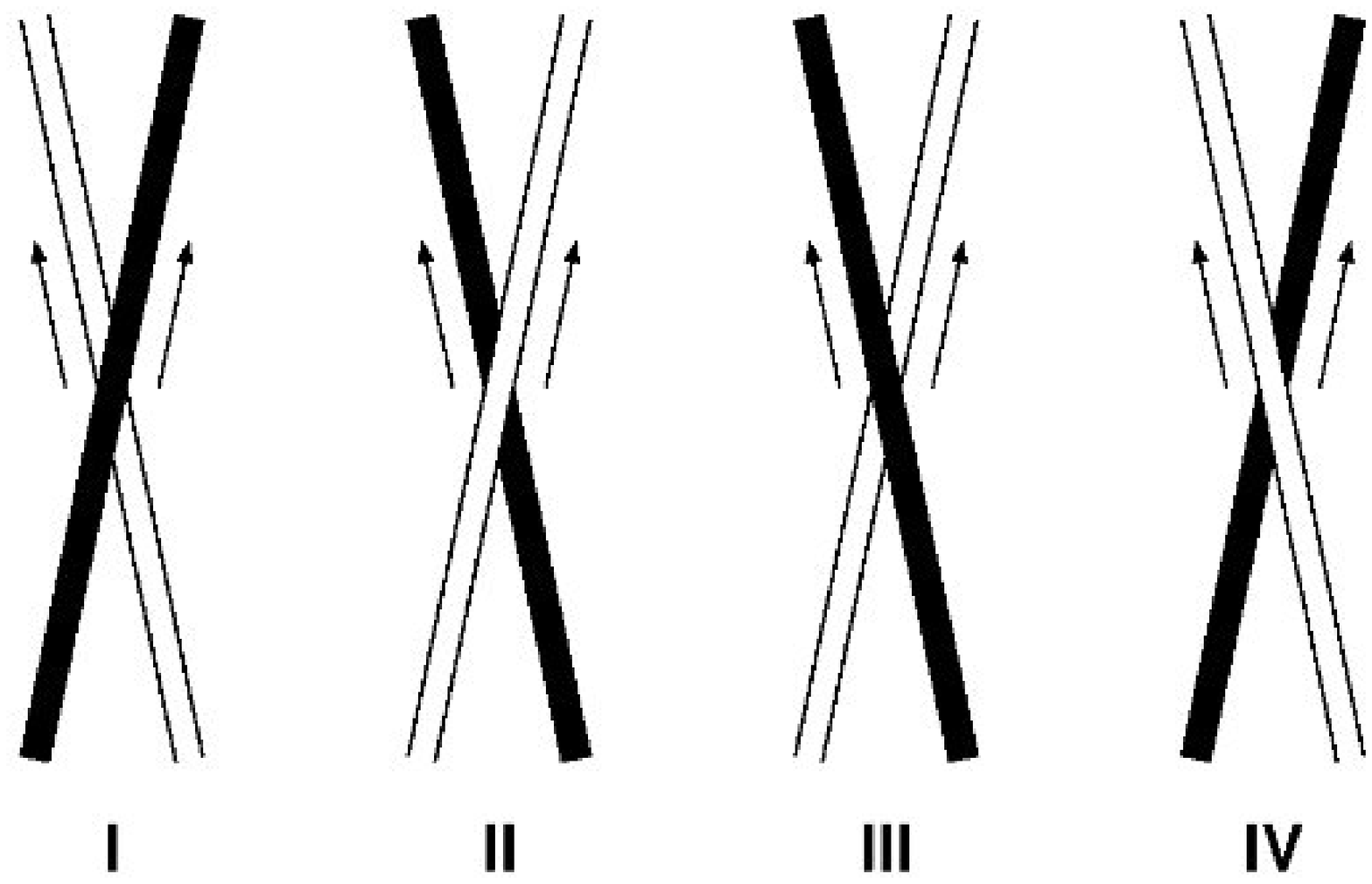}
\includegraphics[height=.25\columnwidth]{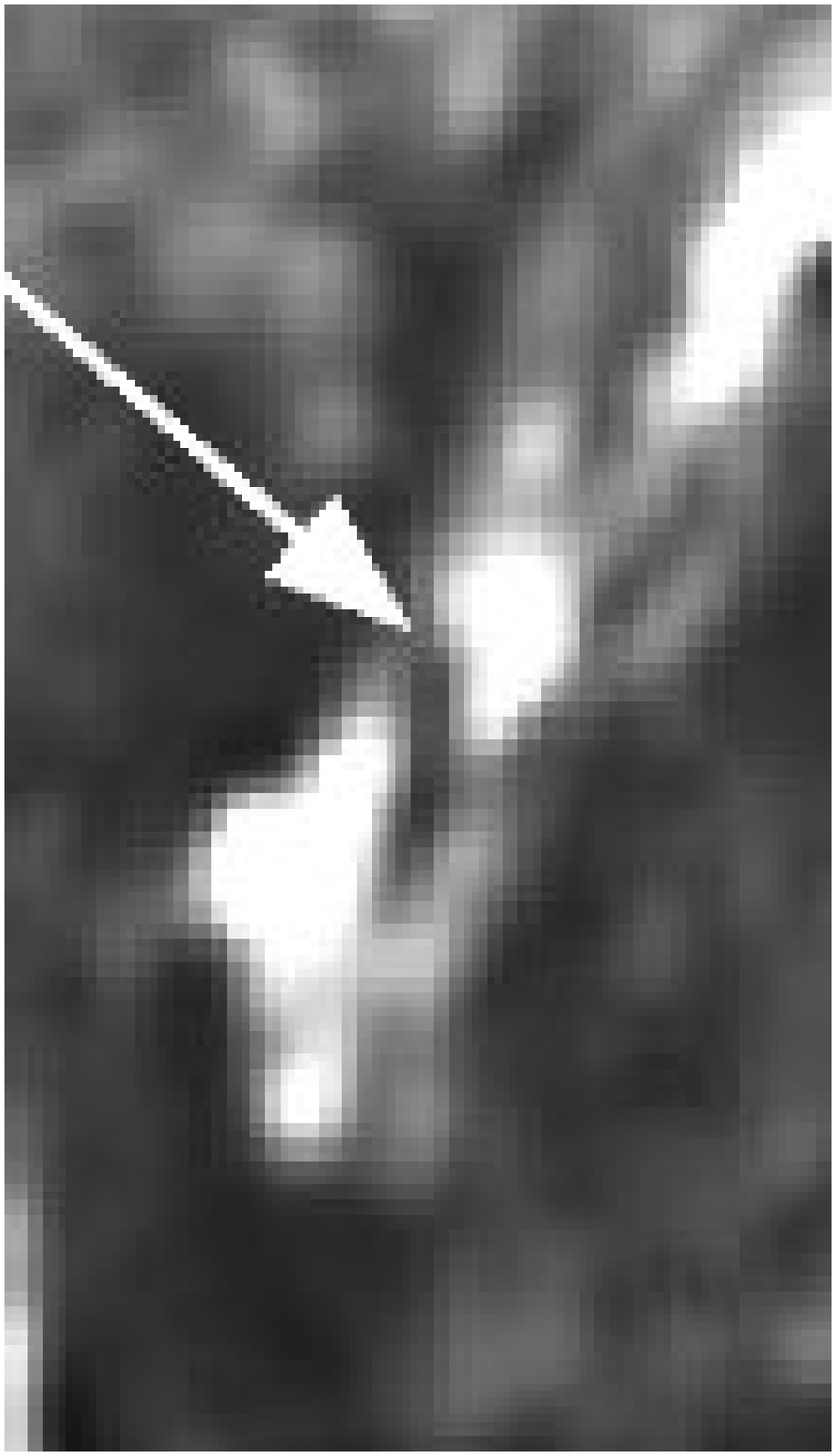}
\end{center}\caption[]{
Sketch of interlocked flux rings of positive helicity with two
right-handed crossings (left), compared with two right-handed mutual
crossings (I and II) and two left-handed mutual crossings (III and IV),
as well as a high-resolution EUV image in 171 Angstroms of a magnetic filament
showing a crossing of type III in active region NOAA 8668 at 23\degr
north on 1999 Aug 17 (right; courtesy of Jongchul Chae).
}\label{interlocked}\end{figure}

Four such pseudoscalars are of particular interest:
mean kinetic helicity density $\bra{\oo\cdot\uu}$, with
$\oo=\nab\times\uu$ being the vorticity,
mean magnetic helicity density $\bra{\AAA\cdot\BB}$, with $\AAA$
being magnetic vector potential such that $\BB=\nab\times\AAA$,
mean current helicity density $\bra{\JJ\cdot\BB}$,
and finally the cross helicity, $\bra{\uu\cdot\BB}$; see
\cite{Yok13} for a recent review, especially on cross helicity.
All these helicities have topological interpretations that refer to the mutual linkage
between interlocked structures, for example two flux rings in the case of
magnetic helicity (\Fig{interlocked}, left), two vortex rings in the case
of kinetic helicity, two current tubes in the case of current helicity,
and a vortex tube with a magnetic flux tube in the case of cross helicity.
This topological interpretation goes back to early work of \cite{Mof69}
and is important for the existence of {\em qualitative} helicity
indicators.

\paragraph{Qualitative helicity indicators.}
Familiar to all of us is the case of kinetic helicity in the Earth's
atmosphere: a glance at weather maps in England and Australia shows
low pressure whirlpools of opposite orientation. This helicity is the direct result
of the Coriolis force acting on large-scale flows.
Likewise the Sun is rotating and has large-scale (supergranulation or larger)
flows which feel the effect of the Coriolis force. Hence on the Sun we see the morphological
or qualitative signatures of helicity in large-scale structures:
H$\alpha$ images reveal {\sf S}-shaped structures
in the south and {\sf N}-shaped structures in the north (i.e., inverted
{\sf S}-shaped structures).
They are referred to as sigmoidal structures and the importance of
interpreting them was recognized by Sara \cite{Mar98,Mar98b}
and many people after her \citep{CHM99,ML01,Gib02}.
These helicity indicators can also be linked to mutual crossings of
magnetic flux structures, which are best seen at EUV wavelengths
where filaments appear occasionally in emission.
This technique, which is due to \cite{Cha00}, requires an additional assumption:
the angle between two adjacent structures is an acute one
(i.e., the invisible arrows of the field vectors point in roughly the same direction)
rather than an obtuse one, as illustrated by the
configurations I--IV in \Fig{interlocked} from his paper.
His study confirms that the magnetic field has negative helicity in the
north (corresponding to crossings of types III and IV with {\sf N}-shaped
or so-called dextral filaments) and positive helicity in the south
(corresponding to crossings of types I and II with {\sf S}-shaped 
or so-called sinistral filaments).
In \Fig{interlocked} we also reproduce an EUV image from his paper
with a filament of type III in the northern hemisphere,
consistent with negative helicity.

\paragraph{Semiquantitative indicators.}
We turn now to semiquantitative indicators, by which we mean quantitative measures
of something which is only qualitatively linked the helicity.
As an important example, we can consider  the product of what is known as the horizontal divergence and horizontal
curl of the velocity,
defined respectively as
\EQA
(\dive\uu)_{\rm h}=u_{x,x}+u_{y,y},\cr
(\curl\uu)_{\rm h}=u_{y,x}-u_{x,y}.
\ENA
Here, commas denote partial differentiation.
The horizontal curl is just the same as the $z$ component
of the usual curl.
The product of  $(\dive)_{\rm h}$ and $(\curl)_{\rm h}$ has been determined for
the Sun using local correlation tracking and local helioseismology by \cite{LGB14}.
In fact, \cite{RBP99} have shown that the product of $(\dive)_{\rm h}$
and $(\curl)_{\rm h}$ is a proxy of kinetic helicity.
This is simply because one of the three terms in $\oo\cdot\uu$ is
$u_z\,(u_{y,x}-u_{x,y})$, but, using the anelastic approximation,
$\nab\cdot(\rho\uu)=0$, where $\rho$ is density, and the definition
of the density scale height,
\EQ
H_\rho=-(d\ln\rho/\dd z)^{-1},
\label{Hrho}
\EN
we have
$u_z=H_\rho\,\dive\uu\approx H_\rho(\dive\uu)_{\rm h}$ and therefore
$\bra{\oo\cdot\uu}\approx3H_\rho\bra{(\dive\uu)_{\rm h}(\curl\uu)_{\rm h}}$.

\paragraph{Quantitative helicity measures.}
The measurement of current helicity density, $\JJ\cdot\BB$, goes back to early
work of \cite{See90}, who determined $B_z$ from circular polarization
measurements, while $B_x$ and $B_y$ (giving $J_z$) were obtained from
linear polarization.
Such measurements are now obtained routinely from solar vector magnetograms.
They all suggest that $\JJ\cdot\BB$ is negative in the north and positive
in the south.
Typical values of $\mu_0\JJ\cdot\BB$ are around $3\G^2\km^{-1}$
\citep{ZBS14}.

Regarding magnetic helicity, there is a notable complication in that
$\bra{\AAA\cdot\BB}$ is {\em gauge-dependent} and changes by adding an
arbitrary gradient term to $\AAA$, which does not change $\BB$.
Exceptions are triply-periodic and infinite domains, for which
$\bra{\AAA\cdot\BB}$ turns out to be gauge-invariant.
However, there exists a quantity called the relative magnetic helicity
\EQ
H_{\rm rel}=\int_V (\AAA+\AAA_{\rm p})\cdot(\BB-\BB_{\rm p})\,\dd V,
\EN
where $\BB_{\rm p}=\nab\times\AAA_{\rm p}$ is a potential field
($\nab\times\BB_{\rm p}=\nullvector$) that satisfies
$\BB_{\rm p}|_{\rm surf}\cdot\nnn=\BB|_{\rm surf}\cdot\nnn$ \citep{BF84}.
It is gauge-independent, but it can only be determined over a finite volume.
There is also a corresponding magnetic helicity flux
$2\oint\EE\times\AAA_{\rm p}$, where $\EE=\JJ/\sigma-\uu\times\BB$ is
the electric field.
Simple examples of magnetic helicity and its flux have been presented by
\cite{BR00} for theoretical models with rigid and differential rotation
as well as with an assumed $\alpha$ effect.
Quantitative measurements for the Sun's magnetic field are in the range
of $10^{46}\Mx^2/\mbox{cycle}$ \citep{DV00,Cha01,WL03}.
This value is easily motivated by standard dynamo theory \citep{Bra09}.

Let us finally comment on the cross helicity density $\uu\cdot\BB$,
or its spatial average $\bra{\uu\cdot\BB}$.
Just like $\bra{\AAA\cdot\BB}$, it is a quantity that is conserved by the
nonlinear interactions of the magnetohydrodynamic equations \citep{Wol58},
but it is often small, i.e., the {\em normalized} cross helicity
$2\bra{\uu\cdot\BB}/\bra{\uu^2+\BB^2}$, which is the ratio of two
conserved quantities (the latter being the total energy), is far
away from its extrema of $-1$ and $+1$, if it was vanishing initially.
An exception is a stratified layer with an aligned magnetic field.
This also applies to the solar wind, where there can be regions where
gravity $\grav$ and the magnetic field are systematically aligned with each other.
In those cases, a finite cross helicity can be driven away from zero.
This can be understood by noting that there are two externally
imposed vectors: gravity with a parallel (vertical) magnetic field $\BB_0$.
The latter is a pseudovector, giving rise to a pseudoscalar $\grav\cdot\BB_0$
that is odd in the magnetic field---just like $\bra{\uu\cdot\BB}$.
Indeed, theoretical and numerical work by \cite{RKB11} showed that
\EQ
\bra{\uu\cdot\bb}=(\tau/3)\,(\urms^2/\cs^2)\,\grav\cdot\BB_0
=-(\etat/H_\rho)\,B_0,
\EN
where $\bb=\BB-\bra{\BB}$ denotes the fluctuating magnetic field, $\tau$
is the correlation time of the turbulence, $\urms$ is its rms velocity,
$\cs$ is the sound speed, $\etat=\tau\urms^2/3$ is the turbulent magnetic
diffusivity, and $H_{\rho}=g/\cs^2$ is the density scale height defined in 
\Eq{Hrho} for an isothermal layer.
Measurements for active regions are in the range $1$--$2\G\km\s^{-1}$
\citep{ZWZ11,RKS12}.

\subsubsection{Large and small length scales}
The length-scale dependence of the different types of helicity can be investigated by looking at the
spectra of magnetic energy and helicity. 

\paragraph{Magnetic helicity spectra.}
Given that the magnetic helicity is a conserved quantity,
it should be zero if it was zero initially.
However, because it is a signed quantity,
``zero'' can consist of a ``mixture'' of pluses and minuses.
These two signs can be segregated spatially (typically into north and
south) as well as spectrally (into large and small scales).
This is discussed in more detail below when we talk about catastrophic
quenching of a large-scale dynamo.
For now it suffices to say that we can define a magnetic helicity spectrum
$\HM(k)$, where $k$ is its wavenumber (inverse length scale), and whose
integral gives the mean magnetic helicity density, i.e.,
\EQ
\int_0^\infty\HM(k)\,\dd k=\bra{\AAA\cdot\BB}.
\EN
If the turbulence or the magnetic field were homogeneous, it can be
related to the Fourier transform of the two-point correlation function
\EQ
M_{ij}(\rr)=\bra{\BB_i(\xx+\rr)\BB_j(\xx)},
\EN
which is independent of $\xx$ owing to the assumption of homogeneity,
and $r=|\rr|$ is the separation.
Its Fourier transform over $\rr$ gives $\tilde{M}_{ij}(k)$, which, under
isotropy, has the representation
\EQ
4\pi k^2\tilde{M}_{ij}(k)=(\delta_{ij}-\hat{k}_i\hat{k}_j)2\mu_0\EM(k)
+\epsilon_{ijk}\ii k_k\HM(k),
\label{Mijk}
\EN
where $\hatkk=\kk/|\kk|$ is the unit vector of $\kk$ and $\EM(k)$
is the magnetic energy spectrum with the normalization
$\int\EM(k)\,\dd k=\bra{\BB^2}/2\mu_0$.
The dependence of $\EM(k)$ and $\HM(k)$ on the modulus $k=|\kk|$
of the wavevector $\kk$ is again an consequence of isotropy.
This relation is slightly modified when applied to the two-dimensional
solar surface, but it allows us to obtain for the first time {\em spectra}
of magnetic helicity, i.e., information about its composition from
different length scales; see \cite{ZBS14} for results that have confirmed
that, in the southern hemisphere, the magnetic helicity is positive on
wavenumbers of about $1\Mm^{-1}$.
This is in agreement with the results presented in \Sec{SurfaceHelicities}.

\begin{figure}[h!]\begin{center}
\includegraphics[width=\columnwidth]{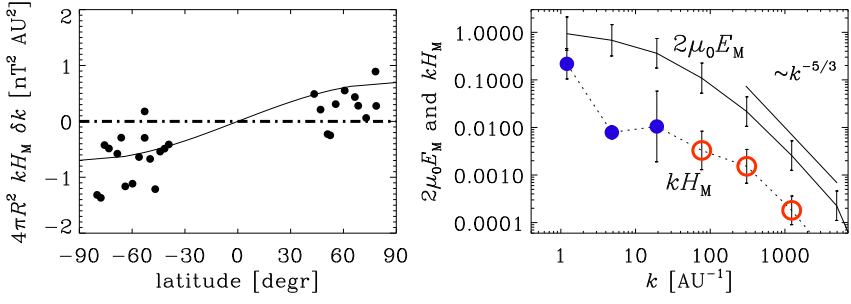}
\end{center}\caption[]{
Latitudinal dependence of spectral magnetic helicity for $k=300\AU^{-1}
\approx2\times10^{-3}\Mm^{-1}$ (left) and the magnetic helicity spectrum
for heliocentric distances above $2.8\AU$ for the northern hemisphere (right).
Filled blue symbols denote negative values and open red ones positive values.
}\label{rmeanhel4b}\end{figure}

\paragraph{Magnetic helicity spectra of the solar wind.}
A similar technique to that presented above has been applied to the
solar wind, where it is possible to obtain the magnetic field vector
from {\it in situ} spacecraft measurements.
The idea to compute the magnetic helicity spectrum by using \Eq{Mijk}
goes back to early work of \cite{MGS82}, who used data from Voyager~1 and 2.
To obtain measurements at positions $\xx$ and $\xx+\rr$, one uses
the Taylor hypothesis to relate the spatial separation $\rr$ to a temporal
separation $t$ through $\rr=\rr_0-\uu_{\rm W} t$, where $\uu_{\rm W}$
is the solar wind velocity of about $800\km\s^{-1}$ and $\rr_0$ is some
reference position.
However, since Voyager~1 and 2 flew close to the equatorial plane, the
resulting magnetic helicity was expected to fluctuate around zero,
which was indeed the case.
This changed when data from {\it Ulysses} were used for such an analysis
\citep{BSBG11}.
In \Fig{rmeanhel4b} we show the resulting spectrum, as well as the latitudinal
dependence a specific $k$.
We should point out that $\HM(k)$ is here gauge-invariant because
we are dealing with an infinite or periodic domain, which is automatically
implied by the use of Fourier spectra.
Note also that $k|\HM(k)| \le 2\mu_0\EM(k)$, which is also known as the
realizability condition.

The importance of magnetic helicity became particularly clear in
connection with understanding the phenomenon of catastrophic quenching,
which will be explained in the next section.

\subsubsection{Magnetic helicity conservation}

To understand the basics of dynamo action, including saturation mechanisms, it 
is often useful to work in idealized periodic geometries, and with turbulence
either from simple forcing or driven by convection.
While early work in this context at small values of $\Rm$ were promising
\citep{BNPST90}, subsequent studies at larger values of $\Rm$ showed that the
field-aligned emf, $\bra{\uu\times\bb}_\parallel/B_\parallel$ approaches zero
as $\Rm\to\infty$ \citep[see][]{CH96}.

This was understood as a consequence of magnetic helicity conservation \citep{GD96}.
Consider the equation for the fluctuation of the magnetic vector
potential $\aaaa=\AAA-\bra{\AAA}$, where angle brackets denote volume averages and lower case symbols the
fluctuations. The mean flow is assumed to vanish.
Thus, we have
\EQ
{\partial\aaaa\over\partial t}=\uu\times\bra{\BB}+\uu\times\bb-\eta\mu_0\jj.
\EN
We now derive the equation for the mean magnetic helicity density
of the small-scale field as
\EQ
{\dd\over\dd t}\bra{\aaaa\cdot\bb}=-2\bra{\uu\times\bb}\cdot\bra{\BB}
-2\eta\mu_0\bra{\jj\cdot\bb}.
\label{ddt_abm1}
\EN
In the steady state, we find that
\EQ
\mbox{``$\alpha$''}\equiv\bra{\uu\times\bb}\cdot\bra{\BB}/\bra{\BB}^2
=-\eta\mu_0\bra{\jj\cdot\bb}/\bra{\BB}^2.
\label{Keinigs}
\EN
This relation for ``$\alpha$'' is sometimes known as Keinigs relation
\citep{Kei83} and shows not only that ``$\alpha$'' is positive when
$\bra{\jj\cdot\bb}$ is negative (i.e., in the north), but also that
``$\alpha$''$\to0$ when $\eta\to0$, i.e., in the limit of large magnetic
Reynolds numbers.
This remarkable result seems like a disappointment to $\alpha$ effect theory,
but it only means that no $\bra{\BB}$, defined as a volume average (!),
can be generated.
This should then be no surprise, because the volume average
of $\BB$ is a conserved quantity for the periodic boundary conditions
used in the study of \cite{CH96} (which is the reason we put $\alpha$
in quotes).

In reality, we are interested in averages that vary in space.

When we consider the case where the averages are allowed to vary in space,
the divergences of magnetic helicity flux will in general be non zero
in the magnetic helicity equation.

\paragraph{Dynamical quenching.}

Consider planar averages, denoted by an overbar, magnetic helicity
conservation yields instead
\EQ
\alpha={\alpK+\Rm\left[\mu_0\meanJJ\cdot\meanBB/\Beq^2
-\nab\cdot\meanFF_{\rm f}/2\Beq^2-(\partial\alpha/\partial t)/(2\etat\kf^2)\right]
\over1+\Rm\meanBB^2/\Beq^2}.
\label{quenching}
\EN
Here, $\meanFF_{\rm f}$ is the magnetic helicity flux from the fluctuating
magnetic field and $\alpK=-(\tau/3)\bra{\oo\cdot\uu}$ is the kinetic
$\alpha$ effect, which itself could depend on $\meanBB$, but this is
here neglected.
The main contribution to the quenching in \Eq{quenching} comes
from the magnetic contribution $\alpM$ to the $\alpha$ effect,
where $\alpha=\alpK+\alpM$ \citep{PFL76}.

\EEq{quenching} confirms first of all that $\alpha$ is catastrophically
quenched (i.e., in an $\Rm$-dependent fashion), when volume averages are
used, i.e., when $\meanJJ=\meanFF_{\rm f}=\nullvector$ and the assumption
of stationarity is made ($\partial\alpha/\partial t=0$).
In that case, we obtain
\EQ
\alpha={\alpK\over1+\Rm\meanBB^2/\Beq^2}
\quad\mbox{(for volume averages!)}.
\label{Volquenching}
\EN
This equation was first motivated by \cite{VC92} on the grounds that
the energy ratio of small-scale to large-scale magnetic fields is
proportional to $\Rm$, i.e., $\bra{\bb^2}/\bra{\meanBB^2}\approx\Rm$
\citep[e.g.][]{Moffatt78,Krause80}, but this relation becomes invalid at
large values of $\Rm$ where the rhs has to be replaced by $\ln\Rm$
\citep{KR94}.
Furthermore, this relation assumes that the small-scale magnetic field
is solely the result of tangling, so dynamo action is actually ignored
in the old argument of \cite{VC92}.

It is now clear from \Eq{quenching} that catastrophic quenching is
alleviated when there are mean currents ($\meanJJ\neq\nullvector$),
which is already the case in triply-periodic helical
turbulence, where, in fact, a super-equipartition field with
$\bra{\meanBB^2}/\bra{\bb^2}\approx\kf/k_1$ can be generated, albeit
only on a resistive time scale \citep{Bra01}.
Here, $\kf/k_1$ is the aforementioned scale separation ratio.
Nevertheless, because of the long saturation time (which is determined by the microphysical diffusivity),
such dynamos cannot be astrophysically relevant.

The ultimate rescue from catastrophic quenching seems to come from the presence
of magnetic helicity flux divergences ($\nab\cdot\meanFF_{\rm f}\neq0$).
Interestingly, what is primarily required is that the dynamo-generated
field is no longer completely homogeneous as in $\alpha^2$ dynamos,
where $\meanBB$ is a Beltrami field with spatially constant $\meanBB^2$.
For example, when there is shear, we can have $\alpha\Omega$-type dynamo
action with finite $\nab\cdot\meanFF_{\rm f}$ within an otherwise periodic
(or shearing-periodic) domain where catastrophic quenching was indeed
found to be alleviated \citep{HB12}.
However, to demonstrate complete $\Rm$ independence is still difficult,
and can only be expected for $\Rm\ga1000$ \citep{DSGB13}.

At this point it is useful to return to the question of gauge-dependence.
The evolution equation for the mean magnetic helicity density of the
fluctuating field, $\overline{\aaaa\cdot\bb}$, can be written with a finite magnetic helicity
flux divergence,
\EQ
{\partial\over\partial t}\overline{\aaaa\cdot\bb}=
-2\,\overline{\uu\times\bb}\cdot\meanBB
-2\eta\mu_0\overline{\jj\cdot\bb}
-\nab\cdot\meanFF_{\rm f}.
\label{ddt_abm2}
\EN
On can now examine whether $\overline{\aaaa\cdot\bb}$, in the gauge
under consideration, happens to be statistically stationary.
In general, this does not need to be the case
\citep[for an example, see Fig.~2 of][]{BDS02}, but if it is,
we can consider the overbars as denoting also an average over time,
because then the left-hand side of \Eq{ddt_abm2} vanishes and we have
\EQ
0=-2\,\overline{\uu\times\bb}\cdot\meanBB
-2\eta\mu_0\overline{\jj\cdot\bb}
-\nab\cdot\meanFF_{\rm f}.
\label{ddt_abm3}
\EN
What is remarkable here is the fact that, at least in this special
case ($\overline{\aaaa\cdot\bb}$ constant in time) the magnetic
helicity flux divergence $\nab\cdot\meanFF_{\rm f}$ is no longer
gauge-dependent, i.e., it must be the same in all gauges.
Moreover, unlike the aforementioned surface-integrated gauge-invariant
magnetic helicity flux $2\oint\EE\times\AAA_{\rm p}$, we can now
make statements about its local dependence and its physical
relation to mean flows and gradients of the magnetic helicity density.
This has been done in several simulations which all confirm
that an important part of the magnetic helicity flux is carried
turbulent-diffusively like in Fickian diffusion \citep{Mit10,HB10,DSGB13}.

\subsubsection{Observational clues}
\begin{figure}[t!]\begin{center}
\includegraphics[height=.25\columnwidth]{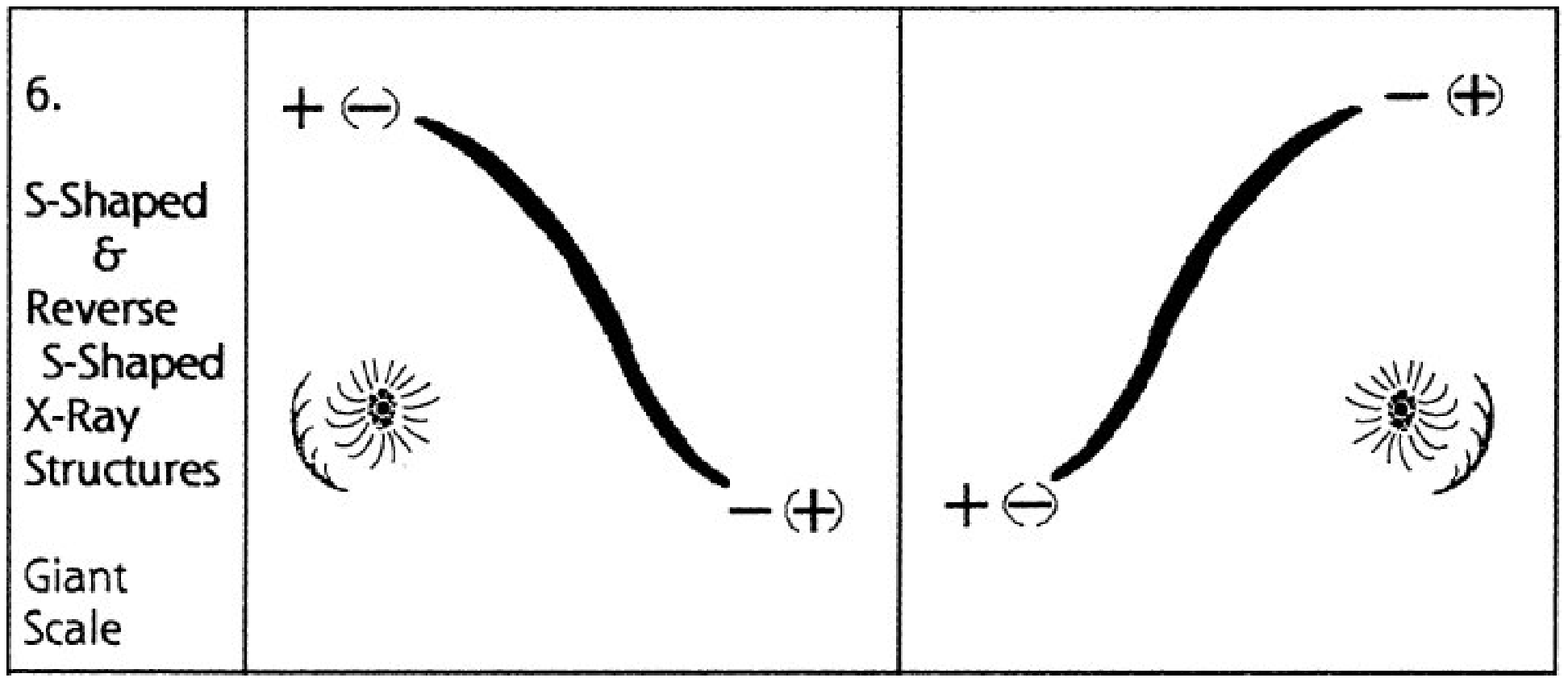}
\includegraphics[height=.25\columnwidth]{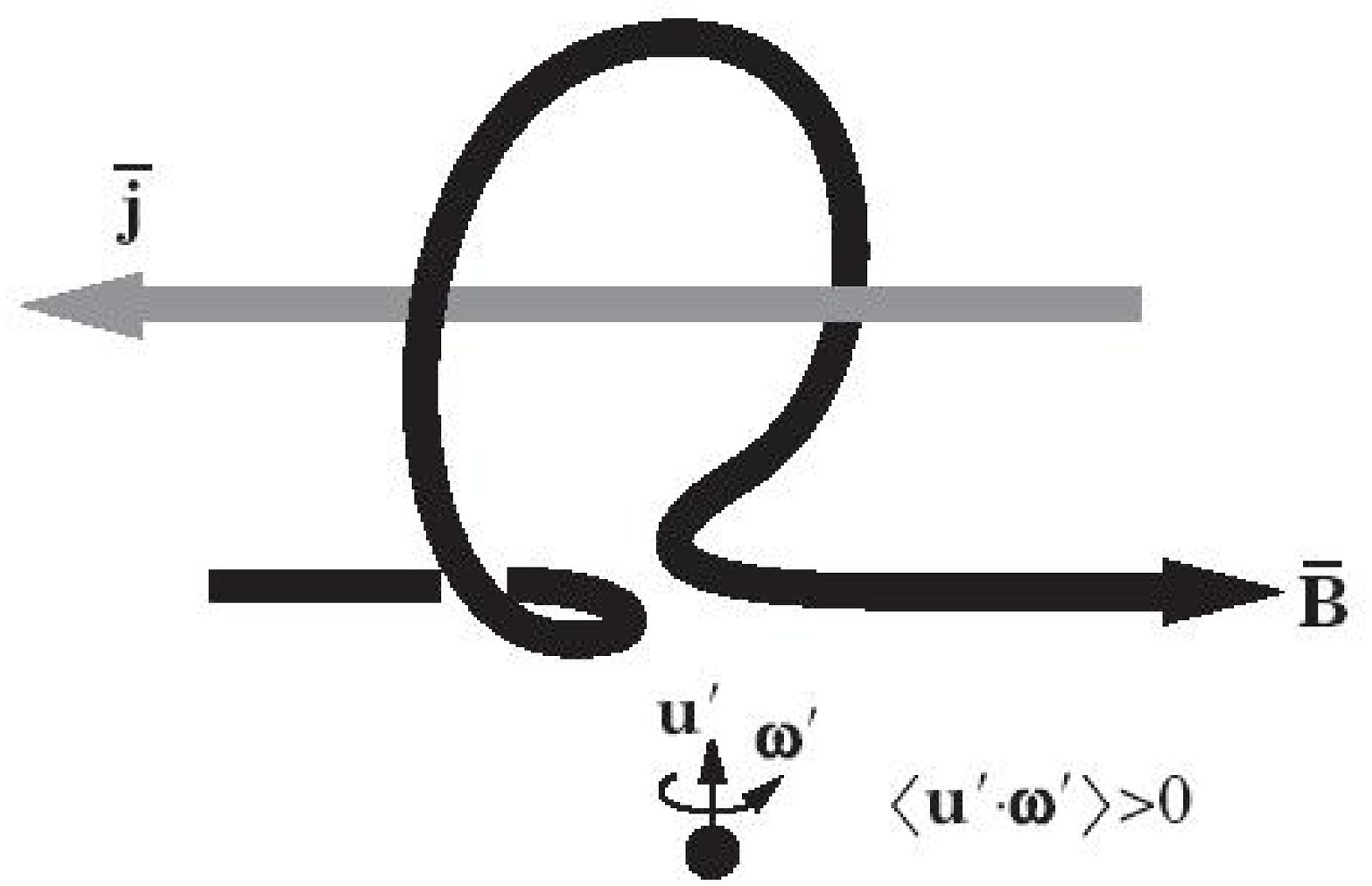}
\end{center}\caption[]{
Sketch of {\sf N}-shaped and {\sf S}-shaped sigmoidal structures
in the northern and southern hemispheres (courtesy of Sara Martin)
and sketch of an $\alpha$ loop for the southern hemisphere with
positive kinetic helicity, taking {\sf S}-shaped form when
viewed from above (courtesy of Nobumitsu Yokoi).
}\label{Martin98}\end{figure}

\paragraph{The sign of the helicity at different spatial scales.}
To make contact with solar magnetic helicity observations, we must
ask about the scales on which helical fields are generated.
If the large-scale field is really generated by an $\alpha$ effect,
then both magnetic and current helicities of the large-scale field
should have the same sign \citep{Bra01,BB02} and should
be positive in the north, where $\alpha>0$.
On large scales, the magnetic helicity $\bra{\meanAA\cdot\meanBB}$ obeys
\EQ
{\dd\over\dd t}\bra{\meanAA\cdot\meanBB}=2\alpha\bra{\meanBB^2}
-2\etaT\mu_0\bra{\meanJJ\cdot\meanBB},
\label{ddt_AbarBbar}
\EN
where we have assumed an isotropic $\alpha$ effect and an isotropic
turbulent magnetic diffusivity $\etat$ and $\etaT=\etat+\eta$ is the
total (turbulent plus microphysical) magnetic diffusivity.
Note that, in the steady-state, we have
\EQ
\mu_0\bra{\meanJJ\cdot\meanBB}=(\alpha/\etaT)\,\bra{\meanBB^2},
\EN
i.e., the magnetic helicity of the large-scale field should
indeed be positive in the north.
At small scales, on the other hand, 
we expect the opposite sign.
Only the latter has been observed directly.
However, the {\sf N}- and {\sf S}-shaped structures in H$\alpha$
images can indirectly be associated with large-scale fields
resulting from a positive $\alpha$ affect in the north;
see \Fig{Martin98}.
Indeed, the barbs of filaments are an example, because right-handed
(left-handed) barbs are found in filaments in which the purely axial
threads (independent of the barb threads) have a slight but definite
shape of a left-handed (right-handed) sigmoid.
This way of interpreting the two signs of helicity within a single filament
was discussed by \cite{Ruzmaikin_etal03}; see Table~1 of \citep{Martin03}
for details.

\paragraph{Helicity reversals within the solar wind.}
The observation of the magnetic helicity spectrum in the
solar wind poses some questions, because the sign is exactly
opposite of what is observed at the solar surface.
Theoretical support for this surprising result comes from simulations
of dynamos with an extended outer layer that only supports turbulent
diffusion, but no $\alpha$ effect.
A magnetic helicity reversal was first recognized in the simulations of
\cite{WBM11,WBM12}, but such reversals were already present in early
mean-field simulations of \cite{BCC09}, which included the physics of
turbulent-diffusive magnetic helicity fluxes.

\begin{figure}[t!]\begin{center}
\includegraphics[height=.267\columnwidth]{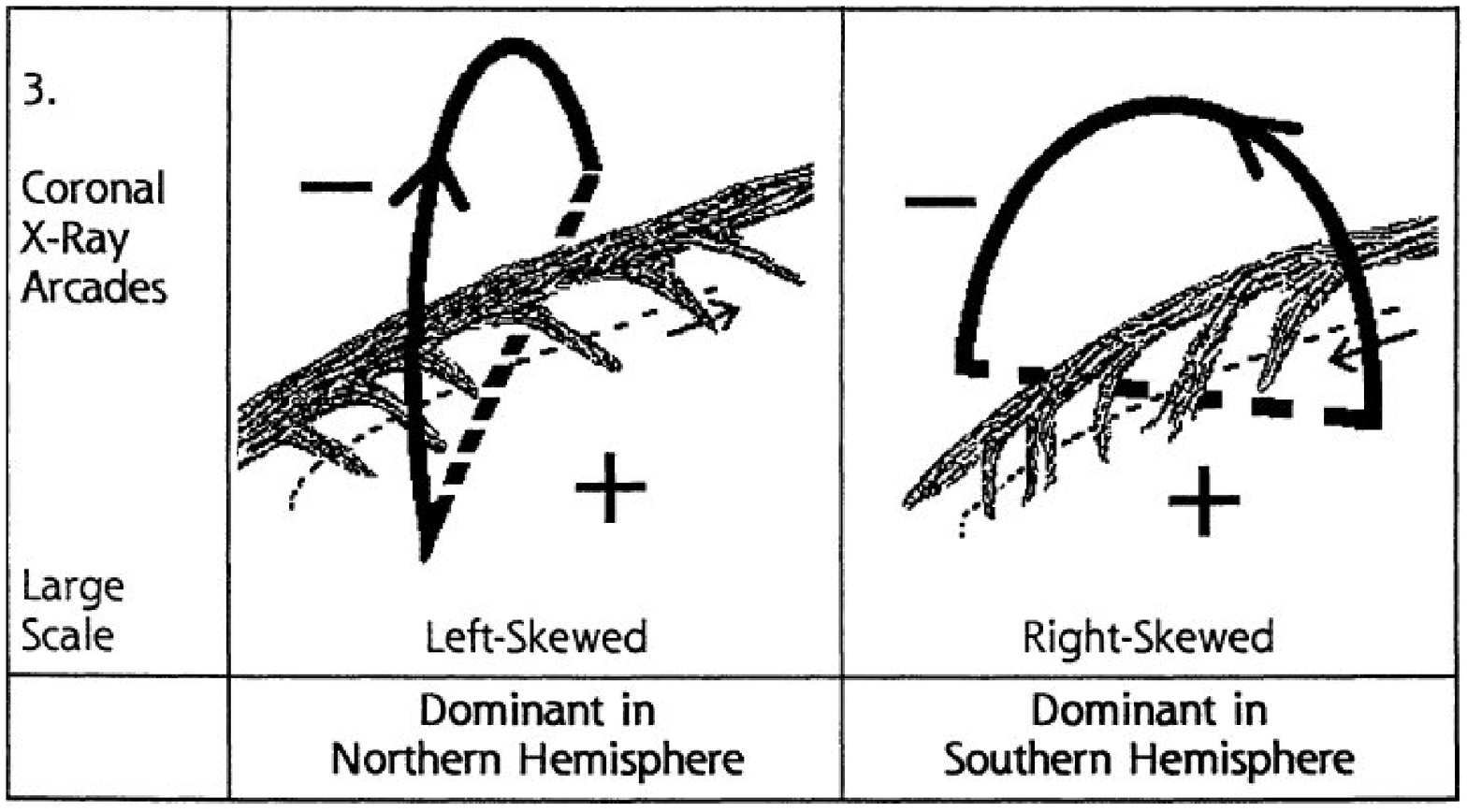}
\includegraphics[height=.273\columnwidth]{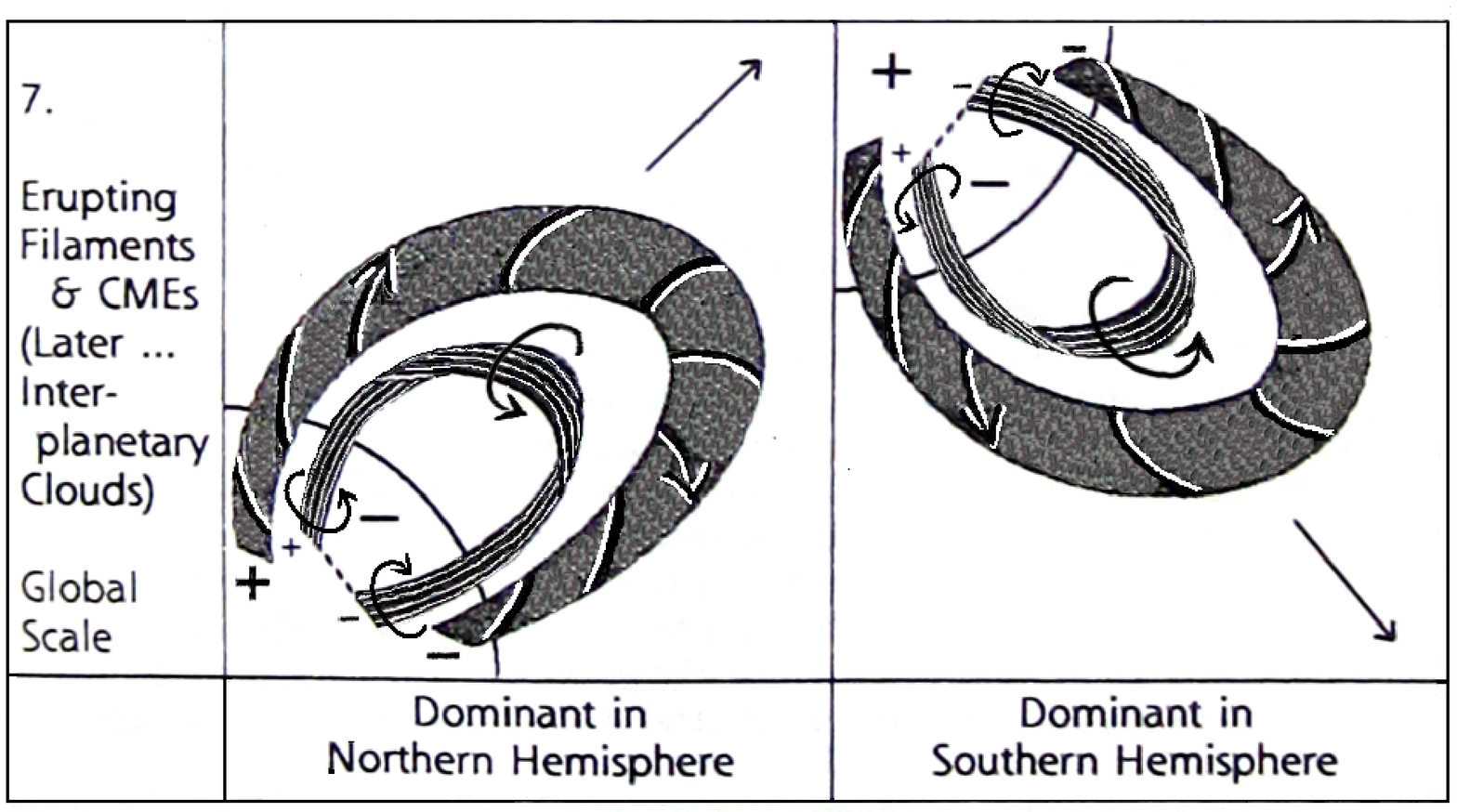}
\end{center}\caption[]{
Sketch of coronal {\sf X}-ray arcades in the northern and southern hemispheres
as well as their respective signs when they turn into interplanetary clouds
(courtesy of Sara Martin).
}\label{Martin98c}\end{figure}

Two related explanations have been proposed.
Firstly, within the dynamo the effects of $\alpha$ and $\etaT$ nearly
balance, which implies that both terms enter the magnetic helicity
equation with opposite signs; see \Eq{ddt_AbarBbar}.
However, within the wind, the $\alpha$ effect is basically absent,
creating therefore an imbalance and thus a contribution of opposite sign.
A related explanation assumes a steady state and invokes a turbulent-diffusive
magnetic helicity flux obeying a Fickian diffusion law, i.e.,
$\meanFF_{\rm f}=-\kappa_{\rm f}\nab\overline{\aaaa\cdot\bb}$,
where $\kappa_{\rm f}$ is a turbulent diffusivity for magnetic
helicity of the small-scale field.
The difference from heat diffusion is that temperature is positive
definite, but magnetic helicity is not.
To transport positive magnetic helicity outward, we need a negative
magnetic helicity gradient, which tends to drive it to (and even through)
zero, which could explain the reversal.
At present it is unclear which, if any, of these proposals is applicable.
It is interesting to note, however, that similar reversals are seen also
in the opposite orientations of coronal {\sf X}-ray arcades
in the northern and southern hemispheres; see \Fig{Martin98c}.
In essence, many filaments develop a sideways rolling motion that begins
from the top down \citep{Martin03} and evidence of this motion was found
in H$\alpha$ Doppler observations and in 304 Angstroms images from SOHO
\citep{PanasencoMartin08} and subsequently in 304 Angstroms from SDO
and STEREO \citep{Panasenco_etal11}. The most convincing evidence that
the forces for this change come from the coronal environment is the
correlation with coronal holes. Quiescent filaments without exception
were found to roll from the top down away from adjacent coronal holes
\citep{Panasenco_etal13}.
The right-hand part of \Fig{Martin98c} shows the direction of motions in
the filament that result in their becoming twisted during eruption.
The general direction of the magnetic field is denoted by the polarity
at the footpoints.

\paragraph{Magnetic structures from cross helicity?}

\begin{figure}[t!]\begin{center}
\includegraphics[width=.98\textwidth]{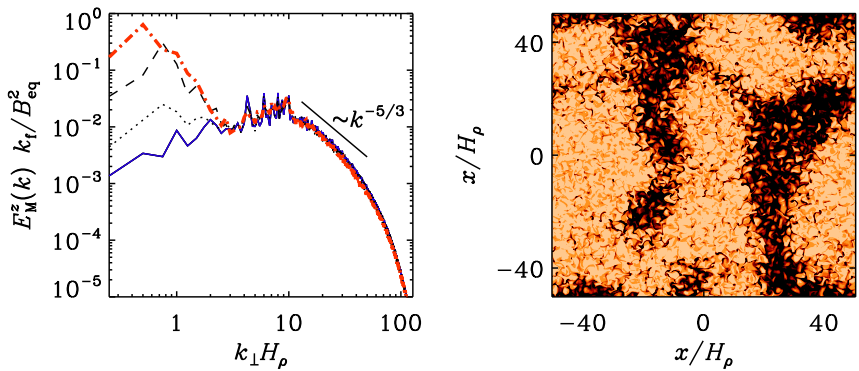}
\end{center}\caption[]{
Left: Normalized spectra of $B_z$ from Run~A40/1 of \cite{BGJKR14}
at turbulent diffusive times $t\etat/H_\rho^2\approx0.2$, 0.5, 1, and 2.7
with $\kf H_\rho=10$ and $k_1 H_\rho=0.25$.
Right: Magnetic field configuration at the upper surface near the end
of the simulation.
}\label{pBz_spec2__pBzm_top_comp_gkf_bern15}\end{figure}

In the presence of strong stratification, cross helicity is being
generated if a large-scale magnetic field pierces the surface.
This leads to a gradual evolution of the magnetic energy spectra
of the vertical field, $E_{\rm M}^z(k)$, showing a growth at small
wavenumbers, akin to inverse transfer resulting from the $\alpha$ effect
and approximate magnetic helicity conservation.
This growth is associated with the development of magnetic structures;
see \Fig{pBz_spec2__pBzm_top_comp_gkf_bern15}.

The formation of such magnetic structures has also been associated with
the possibility of a large-scale instability resulting from a negative
contribution to the effective (turbulent) magnetic pressure.
This instability is therefore referred to as negative effective magnetic
pressure instability (NEMPI).
Earlier work on NEMPI has shown a remarkably degree of predictive power of
this theory in comparison with simulations \citep{BKKMR11,KBKMR13,LBKR13},
but it is still unclear what drives the magnetic structures in simulations
where the field strongly exceeds equipartition with the energy
density of the turbulent motions \citep{MBKR14}.
Whether or not NEMPI or similar phenomena play a role in the formation
of active regions or even sunspots is however an open question.

\section{Conclusion}

The question appears simple: Why does a star such as the Sun support have a global-scale
magnetic field which reverses every 11 years and has equatorially propagating activity wings?
We do not have anything like a complete answer to this question.

At the most fundamental level we do not we do not know the profiles of all the large-scale flows that are present.
Beginning with convection, which is the driver for the global-scale motions and the magnetic field,
we have to leave the reader with the following unanswered questions:
Why is there so little power at scales larger than supergranulation in the solar photosphere \citep{Lord14}?
Does this lack of power reflect a lack of power in these large scales at depth
\citep[compare the results in][]{Hanasoge12, Greer15}? Is our fundamental picture of convection
inapplicable to the solar convection zone \citep{Spr97, Bra15}?

Moving to the flows on the global scale: What does the meridional circulation look like beneath the solar surface
\citep{Zhao13, Schad13, Jackiewicz15}? What maintains the solar differential rotation? What maintains the
meridional circulation? For the latter two questions we have some ideas (which give different answers),
but they depend partly on the strength of the convective flows and will also depend on how the 'small-scale'
turbulent dynamo, supposed to operate in the convection zone, modifies the flows.

These problems in our understanding, for example the power spectrum of the convective flows in the Sun,
are not necessarily critical for understanding the solar dynamo if we take the observed flows as given. 
This simplifies the problem to asking how the (partially) observed flows produce the solar dynamo with all its observed features.
Even this more limited problem is not solved. In this review we have briefly highlighted two important 
open questions. One of these questions is: Why do sunspots only appear at latitudes below about $\pm 40^{\circ}$?
The commonly given answer concerns the stability of toroidal flux located at the base of the convection zone,
which for $10^5$~G flux tubes are much more unstable at low latitudes than at high latitudes. Little consideration
has however been given in this regard to the possibility that the toroidal flux is stored in the bulk of the
convection zone. The second question we have highlighted is what causes the equatorial propagation of the activity belt
towards the equator, and here we have competing answers. Both of these problems appear to be difficult, and their solution
is likely to come from improvements in helioseismology and  modeling efforts.

In order to end on a bright note, we think it is important to point out some of the recent progress which has occurred.
For example we have Hale's law which tells us that the structure of the toroidal
flux is quite simple, and at solar minimum the surface radial field is also simple (it is mainly near the poles
and mainly of one sign in each hemisphere). This simplicity has enabled some rather strong conclusions to be drawn:
the solar dynamo is an alpha-omega dynamo of the Babcock-Leighton type \citep{Cameron15}. Furthermore
we have a number of ideas as to why the amount of activity varies from cycle to cycle \citep[e.g.][]{Jiang15},
and critically we are beginning to test these models against real data \citep{Dikpati06, Choudhuri07, Dikpati10, Nandy11}.
We thus have a long way to go but are making progress and have the right tools and perspective to make
real progress in the coming years.

\begin{acknowledgements}

This work was, in part, carried out in the context of Deutsche 
Forschungsgemeinschaft SFB 963 ``Astrophysical Flow Instabilities and Turbulence''
(Project A16). 

This work was supported in part by
the Swedish Research Council grants No.\ 621-2011-5076 and 2012-5797,
as well as the Research Council of Norway under the FRINATEK grant 231444.

This work has been partially supported by NASA grant NNX08AQ34G.
The National Center for Atmospheric Research is sponsored by the National Science Foundation.

NSO/Kitt Peak data used here are produced cooperatively
by NSF/NOAO, NASA/GSFC, and NOAA/SEL. This work utilizes SOLIS data
obtained by the NSO Integrated Synoptic Program (NISP), managed by the
National Solar Observatory, which is operated by the Association of
Universities for Research in Astronomy (AURA), Inc. under a cooperative
agreement with the National Science Foundation. The synoptic magnetograms
were downloaded from the NSO digital library, http://diglib.nso.edu/ftp.html

We acknowledge the support from ISSI Bern, for her participation in the 
ISSI workshop on.

\end{acknowledgements}

\clearpage
\bibliographystyle{aps-nameyear}  
\bibliography{dynamo}             

\begin{thebibliography}{171}
\bibitem[\protect\citeauthoryear{{Carrington}}{1858}]{Carrington58}
\begin{barticle}
\bauthor{\binits{R.C.} \bsnm{{Carrington}}},
\batitle{{On the Distribution of the Solar Spots in Latitudes since the
  Beginning of the Year 1854, with a Map}}.
\bjtitle{\mnras}
\bvolume{19},
\bfpage{1}--\blpage{3}
(\byear{1858})
\end{barticle}
\endbibitem
\end{thebibliography}

\end{document}